\documentclass{amsart}
\usepackage{graphicx, mathtools, tikz-cd}

\usepackage{amsfonts}
\usepackage{amssymb}
\usepackage{amsthm}
\usepackage{comment}
\usepackage{mathrsfs}
\usepackage{amscd}
\usepackage[all]{xy}
\usepackage{amsbsy}

\usepackage{fullpage}

\usepackage{amsopn}

\usepackage{tikz}
\usetikzlibrary{decorations.footprints}
\usetikzlibrary{decorations.markings}
\usetikzlibrary{decorations.pathmorphing}
\usetikzlibrary{shapes,arrows}

\usepackage{hyperref}

\newtheorem{theorem}{Theorem}[section]

\theoremstyle{definition}
\newtheorem{definition}[theorem]{Definition}

\def\co{\colon\thinspace}

\def\calgd{\mathsf{C}^*}

\def\xto{\xrightarrow}

\DeclareMathOperator{\Sym}{Sym}
\DeclareMathOperator{\csym}{\widehat{\Sym}}

\DeclareMathOperator{\Tr}{Tr}

\DeclareMathOperator{\Maps}{Maps}

\DeclareMathOperator{\MC}{MC}

\DeclareMathOperator{\dgMan}{ndgMan}
\DeclareMathOperator{\gMan}{dgMan}

\DeclareMathOperator{\Man}{Man}
\DeclareMathOperator{\Sets}{Sets}
\DeclareMathOperator{\Grpds}{Groupoids}

\def\fg{\mathfrak g}
\def\fh{\mathfrak h}

\def\ot{\otimes} 

\def\L8{L_\infty}

\def\cinf{C^\infty}

\def\cA{\mathcal A}\def\cB{\mathcal B}
\def\cE{\mathcal E}\def\cF{\mathcal F}\def\cG{\mathcal G}
\def\cI{\mathcal I}\def\cL{\mathcal L}
\def\cM{\mathcal M}\def\cO{\mathcal O}
\def\cS{\mathcal S}
\def\cV{\mathcal V}

\def\CC{\mathbb C}

\def\LL{\mathbb L}
\def\NN{\mathbb N}
\def\RR{\mathbb R}\def\TT{\mathbb T}

\def\ZZ{\mathbb Z}

\def\sE{\mathscr E}
\def\sJ{\mathscr J}
\def\sO{\mathscr O}

\def\bB{\mathbf B}

\def\fJ(E){\mathfrak E}
\def\fJ{\mathfrak J}

\def\fX{\mathfrak X}

\begin{document}
\markboth{Ryan E. Grady}{Quantizing Derived Mapping Stacks}

%
%

\title{Quantizing Derived Mapping Stacks}

\author{Ryan E. Grady}

\address{Department of Mathematical Sciences, Montana State University\\
Bozeman, USA\\
ryan.grady1@montana.edu}

\maketitle


\begin{abstract}
In this review we discuss several topological and geometric invariants obtained by quantizing $\sigma$-models. More precisely, we don't quantize the entire mapping stack of fields, but rather only the substack of low energy fields. The theory restricted to this substack can be presented Lie theoretically and the problem is reduced to perturbative gauge theory. Throughout, we make extensive use of derived symplectic geometry and the BV formalism of Costello and Gwilliam. Finally, we frame the AJ Conjecture in knot theory as a question of quantizing character stacks.

\end{abstract}


\setcounter{tocdepth}{2}
\tableofcontents

\section{Introduction}

Studying the space of maps between geometric objects is commonplace in physical mathematics. Often times this space admits an action functional whose critical points have desired mapping properties, e.g., conformality, holomorphicity, etc. Further, quantizing this system can illustrate surprising topological and geometric properties of the source and/or target object. Hence, $\sigma$-models continue to be a useful tool/study in both physics and mathematics.

In the present work, we discuss the quantization of $\sigma$-models in the Batalin--Vilkovisky (BV) formalism as developed by Costello and Gwilliam.\cite{CG1,CG2} To be more precise, we don't quantize the full space of fields, but rather find an effective quantization of the low energy sector of the theory. This quantization is found by modeling the low energy sector as a gauge theory and then quantizing this gauge theory. The key technical tool in this transformation is that of $\L8$ spaces.

One output of the Costello--Gwilliam approach is that the quantum observables possess significant algebraic structure, that of a {\it factorization algebra}. We present a number of examples of this quantization scheme and discuss the associated algebras of observables. Through this process, we recover deformation quantization, the algebraic index theorem, and various physically relevant vertex algebras.

Next, we extend BV to include manifolds with boundary/boundary conditions. In this case, the boundary observables have the structure of a module over the bulk observables. Specific examples illustrate holography and the CS/WZW correspondence. This setting also yields a conceptual proof that Kontsevich's approach to deformation quantization and Fedosov's approach to deformation quantization are equivalent for symplectic manifolds.

Finally, we outline a derived geometric approach to a conjecture of Garoufalidis\cite{Gar} relating certain knot invariants. The so called {\it AJ conjecture} roughly states that the colored Jones polynomial of a knot determines the knot's $A$-polynomial.

\subsection{Mapping stacks and $\L8$-models}

There are many frameworks in which to consider the space of maps $\Maps (N,M)$ between smooth manifolds. Indeed, one could work in one of several settings of infinite dimensional manifolds; such techniques have been quite successful historically. In the present, we take a more ``generalized"/synthetic/algebro-categorical approach. Indeed, any manifold $M$ determines (via the Yoneda Embedding) a functor
\begin{equation}
\underline{M} : \Man^{op} \to \Sets , \quad N \mapsto \Maps (N,M).
\end{equation}
Though simple, this observation has proved useful, especially its extension to supermanifolds and graded geometry. 

Often there are symmetries for which one would like to account, which leads to the setting of stacks.  A {\it stack}---to first approximation---can be described as a functor $\fX : \Man^{op} \to \Grpds$. If one wishes to allow higher symmetries, then one observes that a groupoid is a particular type of simplicial set and replaces the target category by $s\!\Sets$ to obtain a {\it higher stack}.

Derived geometry has risen to prominence in several areas of physical mathematics. A simple example is given by resolving an intersection, e.g., the derived critical locus of an action functional. In our paradigm, ``deriving" equates to replacing the domain category to allow resolutions of manifolds. Our choice in this case is to consider the category of differential graded manifolds, $\gMan$. A {\it derived stack} is then a functor $\fX : \gMan^{op} \to s\!\Sets$.
We further want any stacky object to preserve weak equivalences in the domain category and to satisfy descent for a given class of covers (equipping $\gMan$ with the structure of a site).

There is an interesting class of derived stacks which are presented by $\L8$ spaces. An $\L8$ space is a pair, $(X, \fg)$, consisting of a smooth manifold and a sheaf of (curved) $\L8$ algebras. Presenting a stack by an $\L8$ space is particularly convenient as computations often reduce to Lie algebra cohomology calculations, for which there are a plethora of algebraic tools.

For present purposes we need to insist on some nilpotent behavior. Let $\dgMan$ be the site of {\it nil dg manifolds}, in which an object $\cM$ is a smooth manifold $M$ equipped with a sheaf $\cO_\cM$ of commutative dg algebras over $\Omega^*_M$ that has a nil dg ideal $\cI_\cM$ such that $\cO_\cM/\cI_\cM \cong \cinf_M$.

An $\L8$ space $B \fg = (X, \fg)$ has an associated ``functor of points'' and hence can be understood as presenting a kind of space $\bB \fg$ in the same way that a commutative algebra presents a scheme. More precisely, to $(X, \fg)$ we associate a simplicial set valued functor
\begin{equation}
\bB \fg: \dgMan^{op} \to s\!\Sets.
\end{equation}
 Moreover, the functor $\bB \fg$ preserves weak equivalences and satisfies \v{C}ech descent, so it defines a derived stack. See Section \ref{sect:Loo} below for further discussion.

\subsection{$\sigma$-Models}

The AKSZ construction of Alexandrov, Kontsevich, Schwarz, and Zaboronsky\cite{AKSZ} is a clever mechanism for producing classical BV theories from topological $\sigma$-models. The inputs into AKSZ are a $k$-shifted symplectic dg manifold and an oriented $(k+1)$-manifold. Roytenberg framed the Courant $\sigma$-model in this framework and classified (a restricted class of) shifted symplectic dg manifolds in low degree.\cite{Roy1,Roy2} Higher Courant $\sigma$-models, i.e., $k\ge 3$, have appeared in the work of Szabo and collaborators.\cite{Szabo}

Pantev, To\"{e}n, Vaqui\'{e}, and Vezzosi\cite{PTVV} extended the AKSZ formalism to the setting of derived algebraic geometry. So the PTVV construction constructs classical action functionals on spaces of maps between oriented and shifted symplectic derived stacks.

Costello\cite{Cos1} has a deformation theoretic to quantizing classical BV theories and his work with Gwilliam provides a general mathematical framework for (perturbative) classical and quantum BV theories. Given a $\sigma$-model, we model the substack of low energy fields inside the stack of all fields by an $\L8$ space. The AKSZ/PTVV  type theory is thereby transformed into a gauge theory (still in the BV formalism). Such a gauge theory is particularly amenable to Costello's techniques as obstruction/deformation computations are reduced to calculations in Lie algebra cohomology.  We highlight a number of examples of the $\sigma$-model/gauge theory transformation in Section \ref{sect:examples}. Moreover, we hope that the Costello--Gwilliam formalism might prove useful in new directions, e.g., \v{S}evera and collaborators' work on generalized Ricci flow in the Courant $\sigma$-model.\cite{SV1, SV2, PSY}


\section{Mapping Stacks and Geometric Structures}

We begin in earnest by recalling shifted symplectic structures on (derived) stacks.  In order to make sense of such symplectic structures, we need to have a notion of closed forms and non-degeneracy. As we see in the next section, in homotopy coherent derived geometry being closed is actually data, not a property.

\subsection{Derived symplectic geometry}

Our focus on derived symplectic geometry is driven by two priorities. The first is to study mapping stacks in the derived setting which often transforms subtle infinite dimensional analytic issues into more algebraic questions. The second is that the BV formalism is intimately related to (-1)-shifted symplectic geometry. Indeed, elementary examples of BV algebras (discussed further below) come from graded finite dimensional vector spaces with a symplectic pairing of degree -1. 

One way to obtain such a (-1)-shifted symplectic vector space is as a shifted cotangent bundle. For simplicity, let $V$ be a finite dimensional vector space, then its shifted cotangent bundle $T^\ast[-1]V= V \oplus V^\vee[-1]$ has a natural symplectic pairing of degree -1 given by the evaluation pairing.  This is the prototype of a BV theory.\cite{Cos1} In general, for any reasonable space/stack $X$, the shifted cotangent bundle $T^\ast[-k]X$ has a degree $-k$ symplectic structure. The adjective ``reasonable" here has many different interpretations\cite{GR,Toen,CS}, which are all unified by the existence of tangent and cotangent complexes and a well behaved notion of closed differential form.

As for a well behaved notion of closed form, we desire a few outcomes:
\begin{itemize}
\item[(i)] The de Rham complex and the complex of closed forms are homotopically coherent. In particular, whether a form is closed does not depend on the {\it presentation} of a stack.
\item[(ii)] The de Rham complex should satisfy (smooth) descent.
\item[(iii)] For $G$ a reductive group, the closed 2-forms of degree 2 on $BG=\ast/G$ are precisely $G$-invariant symmetric bilinear forms, i.e, $Sym^2 (\fg^\vee)^G$.
\end{itemize}
Item (iii) is in some sense a normalization/sanity check. One take-away is that being closed is structure/data as opposed to a property of a form.\cite{Damien18, PTVV} 

Let $X$ be a (derived) stack with cotangent complex $\LL_X$. We will assume that $\LL_X$ is dualizable and the dual will be the tangent complex $\TT_X$. In the setting of dg manifolds, $\LL_X$ and $\TT_X$ are the usual (dg) cotangent bundles and tangent bundles. Define the complex of closed 2-forms, $\Omega^{2,cl} (X)$ as
\begin{small}
\begin{equation}
\Omega^{2,cl} (X) := \mathrm{Tot} \left ( \Gamma \left (X, \Sym^2  (\LL_X[-1]) \right) \xrightarrow{d_{dR}} \Gamma \left(X, \Sym^3 (\LL_X[-1] ) \right ) \xrightarrow{d_{dR}}  \dotsb \right ).
\end{equation}
\end{small}
Let $i: \Omega^{2,cl} (X) \to \Omega^2 (X)$ be the brutal truncation at the first stage.

\begin{definition}
An  $n$-shifted symplectic form on $X$ is a closed 2-form $\omega$ of cohomological degree $n$ such that the induced map $i(\omega)^\sharp : \TT_X \to \LL_X[n]$ is an equivalence.
\end{definition}

Explicitly, a closed 2-form $\omega$ of degree $n$ is a sequence $(\omega_0, \omega_1, \dotsc )$ with $\omega_i \in \Omega^{2+i} (X)$ of internal degree $n-i$ such that $d_{dR} \omega_i = \partial \omega_{i+1}$, where $\partial$ is the internal differential. Non-degeneracy is only a property of $\omega_0 = i (\omega)$, while being closed is a lift of $\omega_0$ to a cocycle in $\Omega^{2,cl}(X)$.

Roytenberg\cite{Roy1} classified symplectic structures on dg manifolds, though his were a special case of the more general definition just given.  A Roytenberg structure is concentrated in a single bidgree and so corresponds to a closed 2-form of type $(\omega_0, 0, 0, 0, \dotsc)$. We revisit Roytenberg's work below when we discuss symplectic structures on Lie algebroids.


\subsection{The $\L8$-model}\label{sect:Loo}

As mentioned above, we take a generalized/functor of points perspective towards geometry. Our goal in this section is to describe how certain mapping stacks can be presented by a nilpotent $\L8$ algebra (or more generally an $\L8$ space). 

More specifically, given a nilpotent $\L8$ algebra, we obtain a functor\cite{Getzler}
\begin{equation}
\bB \fg: \dgMan^{op} \to s\!\Sets, \quad (M, \cI_M, \cO_M) \mapsto \MC_\bullet(\fg \otimes \cI_M).
\end{equation}
For any nilpotent $\L8$ algebra $\fh$, the space $MC_\bullet (\fh)$ is the \emph{Maurer--Cartan} space, explicitly an $n$-simplex is an element $\alpha \in \fh \otimes \Omega^\ast (\Delta^n)$ satisfying the Maurer--Cartan equation; the simplicial set structure is inherited from the face and degeneracy maps between geometric simplexes as $n$ varies. The functor $\bB \fg$ preserves (and detects) weak equivalences and satisfies \v{C}ech descent on the site of nilpotent dg manifolds.\cite{GGLoop}

The functor $\bB \fg$ can be adapted to the setting of $\L8$ spaces, i.e., a manifold $X$ equipped with a sheaf of (curved) $\L8$ algebras $\fg_X$, see below for a precise definition. Given a smooth map $f: M \to X$, we obtain a curved $\L8$ algebra over $\Omega^\ast_M$ which we denote $f^\ast \fg_X$. An $n$-simplex in $\bB \fg_X (M, \cI_M, \cO_M)$ is then a pair $(f, \alpha)$ where $f: M \to X$ is smooth and $\alpha \in \MC_n \left (f^\ast \fg_X \otimes_{\Omega^\ast_M} \cI_M \right )$. The functor $\bB \fg_X$ is still a derived stack and by constructing an appropriate $\fg_X$, we can represent many derived mapping stacks.

\subsubsection{Conventions}

We work throughout in characteristic zero and with cohomologically, so the differential in any complex increases degree by one.

For $A$ a cochain complex, $A^\sharp$ denotes the underlying graded vector space. If $A$ is a cochain complex whose degree $k$ space is $A^k$, then $A[1]$ is the cochain complex where $A[1]^k = A^{k+1}$. We use $A^\vee$ to denote the graded dual.

For $V$ a graded $R$-module, its {\em completed symmetric algebra} is the graded $R$-module
\begin{equation}
\csym_{R} (V) = \prod_{n \geq 0} \Sym^n_{R}(V)
\end{equation}
equipped with the filtration $F^k \csym_{R} (V)  = \Sym^{\geq k}_{R}(V)$ and the usual commutative product, which is filtration-preserving. 

\subsubsection{$\L8$ algebras and spaces}

Note that we allow nonzero {\it curving} in our algebras; the geometric examples of interest are naturally curved, see also \cite{GGLoop} for further interpretative remarks.

\begin{definition}
Let $A$ be a commutative dg algebra with a nilpotent dg ideal $I$. A {\em curved $\L8$ algebra over $A$} consists of
\begin{enumerate}
\item[(i)] a locally free, $\ZZ$-graded $A^\sharp$-module $V$, and
\item[(ii)] a linear map of cohomological degree 1, $d: \Sym (V[1]) \to  \Sym (V[1])$,
\end{enumerate}
where $\Sym (V[1])$ indicates the graded vector space given by the symmetric algebra over the graded algebra $A^\sharp$ underlying the dg algebra $A$. Further, we require 
\begin{enumerate}
\item[(a)] $d^2 = 0$,
\item[(b)] $(\Sym (V[1]),d)$ is a cocommutative dg coalgebra over $A$ (i.e., $d$ is a coderivation), and
\item[(c)] modulo $I$, the coderivation $d$ vanishes on the constants (i.e., on $\Sym^0$).
\end{enumerate}
\end{definition}

We use $C_\ast(V)$ to denote the cocommutative dg coalgebra $(\Sym (V[1]),d)$.  There is also the natural Chevalley--Eilenberg \emph{cohomology} complex $C^\ast(V) := (\csym (V^\vee[-1]),d)$, where the notation $\csym (V^\vee[-1])$ indicates the completed symmetric algebra over the graded algebra $A^\sharp$ underlying the dg algebra $A$.  The differential $d$ is the ``dual" differential to that on $C_*(V)$. In particular, it makes $C^*(V)$ into a commutative dg algebra, so $d$ is a derivation.

The $n$-fold brackets of $V$ are obtained from $d$ as follows. A derivation is determined by its behavior on $V^\vee[-1]$, thanks to the Leibniz rule. Hence we may view $d$ as simply an $R$-linear map from $V^\vee[-1]$ to $\csym(V^\vee[-1])$. Consider the homogeneous components of $d$, namely the maps $d_n: V^\vee[-1] \to \Sym^n(V^\vee[-1])$. If we dualize, we get maps
\begin{equation}
\ell_n : \Sym^n(V^\vee[-1])^\vee \to (V^\vee[-1])^\vee,
\end{equation}
which we can consider as degree $0$ maps from $(\wedge^n V)[n-2]$ to $V$. These are the Lie brackets on $V$, and we sometimes call them the {\it Taylor coefficients} of the bracket. The higher Jacobi relations between the $\ell_n$ are encoded by the fact that $d^2 = 0$.

Also relevant are families of curved $\L8$ algebras parametrized by a manifold $X$.

\begin{definition}
Let $X$ be a smooth manifold. An {\it $\L8$ space} is a pair $(X, \fg)$, where $\fg$ is the sheaf of smooth sections of a $\ZZ$-graded vector bundle $\pi: V \to X$ equipped with the structure of a curved $\L8$ algebra structure over the commutative dg algebra $\Omega^\ast_X$ with nilpotent ideal $\cI = \Omega^{\geq 1}_X$.
\end{definition}

For brevity, we sometimes write $B \fg$ for the $L_\infty$ space $(X, \fg)$ (see the notation below).  By definition, functions on $B \fg$ are given by $\cO (B \fg) = C^\ast (\fg)$.

\subsubsection{Examples}

The following are $\L8$ spaces which encode various geometric structures on manifolds. All these examples arise via a Fedosov resolution process. In what follows, for $E$ a vector bundle, $J(E)$ will denote the associated infinite jet bundle. Similarly, given a $D$-module---such as $J(E)$)---$dR(-)$ will denote the associated de Rham complex.

\begin{itemize}
\item[(i)] Consider the $L_\infty$ space $(X, 0)$, for $X$ a smooth manifold. This $L_\infty$ space presents a version of the {\it de Rham stack} $X_{dR}$.  For any dg manifold $(M, \cO_\cM)$, we have $X_{dR} (M, \cO_\cM) = X_{dR} (M, \cinf_M)$, which is the constant simplicial set of smooth maps $M \to X$.
\item[(ii)] Let $X$ be a smooth manifold. There is an $L_\infty$ space $(X, \fg_X)$ such that
\begin{itemize}
\item $\fg_X \cong \Omega^\sharp_X(T_X[-1])$ as  $\Omega^\sharp_X$ modules, and
\item $C^*(\fg_X) \cong dR(\sJ) $ as commutative $\Omega_X$ algebras, where $\sJ$ is the infinite jet bundle.
\end{itemize}
$\bB \fg_X$ is a natural {\it derived enhancement} of the smooth manifold $X$,\cite{GGLoop} and we will use it below to describe an explicit model for the {\it derived loop space}.
\item[(iii)] Let $Y$ be a complex manifold, then there exists an $\L8$ space $(Y, \fg_{Y_{\overline{\partial}}})$, such that
\begin{itemize}
\item As an $\Omega_Y^\sharp$-module, $\fg_{Y_{\overline{\partial}}}$ is isomorphic to $\Omega^\sharp_Y(T^{1,0}_Y [-1])$;
\item The derived stack $\bB \fg_{Y_{\overline{\partial}}}$ represents the moduli problem of holomorphic maps into $Y$.
\end{itemize}
\item[(iv)] Generalizing the previous constructions, in Ref. \cite{GGAlgd} we associate an $L_\infty$ space to any Lie algebroid.  That is, let $\rho \co L \to T_X$ be a Lie algebroid over a smooth manifold $X$, then there exists an $\L8$ space $(X, \fg_L)$
such that
 \begin{itemize}
 \item $\fg_L \cong \Omega^\sharp_X (T_X[-1] \oplus L)$ as $\Omega^\sharp_X$ modules, and
 \item $C^* ( \fg_L) \cong dR(J(\calgd(L)))$ as commutative $\Omega^\ast_X$ algebras.
 \end{itemize}
 \item[(v)] Let $(X, \Pi)$ be a Poisson manifold, then by the previous example there is a $\L8$ space, $(X, \fg_\Pi)$, associated to the Lie algebroid $T^\vee_X \xrightarrow{\Pi^\flat} T_X$.
\end{itemize}

Let us articulate the second statement of example (iii) more precisely. Costello \cite{CosWG2} proves that for any complex manifold $Z$ (viewed as the nilpotent dg manifold $(Z, \Omega^{0, \ge 1}_Z , \Omega^{0,\ast}_Z)$), $\bB \fg_{Y_{\overline{\partial}}}(Z)$ is the discrete simplicial set of holomorphic maps from $Z$ to $Y$, i.e., all higher simplices are constant and the zero simplices are in bijection with holomorphic maps $Z \to Y$.

\subsubsection{Shifted symplectic structures on $L_\infty$ spaces}

Recall from above that in order to define symplectic structures, we need to define (co)-tangent objects, and (closed) forms.

\begin{definition}\label{defn:vb}
Let $(X, \fg)$ be an $\L8$ space.  A {\it vector bundle} on $(X,\fg)$ is a $\ZZ$-graded vector bundle $\pi:V \to X$ where the sheaf of smooth sections $\cV$ over $X$ is equipped with the structure of an $\Omega^\sharp_X$-module and where the direct sum of sheaves $\fg \oplus \cV$ is equipped with the structure of a curved $\L8$ algebra over $\Omega^*_X$, which we denote $\fg \ltimes \cV$, such that
\begin{itemize}
\item the maps of sheaves given by inclusion $\fg \hookrightarrow  \fg \ltimes \cV$ and by the projection $\fg \ltimes \cV \to \fg$ are maps of $\L8$ algebras, and
\item the Taylor coefficients $\ell_n$ of the $\L8$ structure vanish on tensors containing two or more sections of $\cV$.
\end{itemize}
The {\it sheaf of sections of $\cV$ over $(X,\fg)$} denotes $C^* (\fg, \cV[1])$, the sheaf on $X$ of dg $C^*(\fg)$-modules given by the Chevalley--Eilenberg complex of $\cV$ as a $\fg$-module. The {\it total space} for the vector bundle $\cV$ over $(X,\fg)$ is the $\L8$ space $(X, \fg \ltimes \cV)$.
\end{definition}

For example, the tangent bundle to $(X, \fg)$ is given by $\fg[1]$ equipped with the adjoint action of $\fg$. Dually, the cotangent bundle is given by $\fg^\vee [-1]$ equipped with the coadjoint action.  It follows that the $k$-forms on $(X, \fg)$ are given by 
\begin{equation}
\Omega^k_{(X,\fg)} = C^* (\fg , (\Lambda^k \fg) [-k]),
\end{equation}
as discussed in \cite{GGLoop}.

Our previous discussion of shifted symplectic structures now applies. In particular, let $\Omega^{2,cl}_{(X,\fg)}$, the complex of {\em closed 2-forms} on the $\L8$ space, be the totalization of the double complex
\begin{equation}
\Omega^2_{(X,\fg)} \xto{d_{dR}} \Omega^3_{(X,\fg)} \xto{d_{dR}} \Omega^4_{(X,\fg)} \xto{d_{dR}} \cdots.
\end{equation}
A {\em closed 2-form} is a cocycle in this complex. Every element $\omega$ of $\Omega^{2,cl}_{(X,\fg)}$ has an underlying 2-form $i(\omega)$ by taking its image under the truncation map $i \co \Omega^{2,cl}_{(X,\fg)} \to \Omega^2_{(X,\fg)}$.
An {\em $n$-shifted symplectic form} on an $\L8$ space $(X,\fg)$ is a closed 2-form $\omega$ of cohomological degree $n$ such that the induced map $i(\omega) \co T_{(X,\fg)} \to T^*_{(X,\fg)}[-n]$ is a quasi-isomorphism.

\subsection{An extended example: the derived loop space}

In derived geometry, there are several flavors of circle, hence there are several loop spaces. We present a few of these loop spaces as $L_\infty$ spaces. For further discussion, see \cite{GGLoop}, \cite{BZN}, or \cite{TV11}.

Recall from above, that a smooth manifold $X$ can be enhanced to a derived stack $\bB \fg_X$ via a Fedosov type resolution. Explicitly, for $X$  a smooth manifold, there is a curved $\L8$ algebra $\fg_X$ over $\Omega_X$, with nilpotent ideal $\Omega^{> 0}_X$, such that
\begin{enumerate} 

\item[(a)] $\fg_X \cong \Omega^\sharp_X(T_X[-1])$ as an $\Omega^\sharp_X$ module;

\item[(b)] $dR(\sJ) \cong C^*(\fg_X)$ as commutative $\Omega_X$ algebras;

\item[(c)] The map sending a smooth function to its $\infty$-jet 
\begin{equation}
C^\infty_X \hookrightarrow dR(\sJ) \cong C^*(\fg_X)
\end{equation}
is a quasi-isomorphism of $\Omega_X$-algebras.

\end{enumerate}

Hence, we can define a derived enhancement of the smooth loop space as follows
\begin{equation}
\cL_{sm}X:  \dgMan^{op}  \to  s\!\Sets, \quad 
 \cM  \mapsto   \bB\fg_X(S^1 \times \cM).
\end{equation}
This space is a derived stack, but it doesn't have a presentation in terms of $L_\infty$ spaces.
Next, consider the $L_\infty$ space $(X, \RR[\epsilon] \otimes \fg_X)$ for $\epsilon$ a square zero parameter of degree 1.  We call this space the {\it Betti loop space} of $X$ and denote it $\cL_{\cB} X$. We have an isomorphism of $L_\infty$ spaces $\cL_{\cB} X \cong T[-1] \bB \fg_X$. The Betti loop space should be thought of as the mapping object obtained by replacing $S^1$ by its cohomology ring.

The final version of the circle we consider is $S^1_{dR} = (S^1 , \Omega^\ast (S^1))$.
This flavor of $S^1$ gives us the {\it de Rham loop space}  of $X$:
\begin{equation}
\cL_{dR}X:  \dgMan^{op}  \to  s\!\Sets, \quad
 \cM  \mapsto   \bB\fg_X(S^1_{dR} \times \cM).
\end{equation}
The de Rham loop space again doesn't have a presentation in terms of an $L_\infty$ space, however a certain substack does.  Consider the substack $\widehat{\cL_{dR} X}$, presented by the $L_\infty$ space $(X, \Omega^\ast (S^1) \otimes \fg_X)$.
\begin{itemize}
\item The derived stack presented by $\widehat{\cL_{dR} X}$ is the substack of $\cL_{dR}X$ where for any dg manifold $(M, \sO_M)$ the underlying map of smooth manifolds $S^1 \times M \to X$ is constant along $S^1$.
\item Any volume form $\omega$ on $S^1$ determines a weak equivalence of derived stacks
$\underline{\omega} :  \cL_{\cB} X \Rightarrow \widehat{\cL_{dR} X}.$
\end{itemize}

Finally, if $X$ is a symplectic manifold, then $\widehat{\cL_{dR}X}$ is a -1-symplectic derived stack. Explicitly,  fix a symplectic form $\omega \in \Omega^2(X)$ and a 1-form $\nu \in \Omega^1(S^1)$ that is not exact, then consider the pairing
\begin{equation}
\begin{array}{rrcl}
\Omega_{\omega,\nu}:&[\fg_X \ot \Omega^*(S^1)]^{\ot 2} &\to &\Omega^*(X)\\[1ex]
&(Z \ot \alpha) \ot (Z' \ot \alpha') &\mapsto &{\displaystyle \int_{\theta \in S^1} }J(\omega)(Z \ot \alpha(\theta), Z' \ot \alpha'(\theta)) \wedge \nu(\theta)
\end{array}
\end{equation}
where $J(\omega)$ is the jet expansion of the symplectic form. The 2-form $\Omega_{\omega, \nu}$ is a -1-symplectic form on $\widehat{\cL_{dR}X}$.\cite{GGLoop}


\section{The Costello--Gwilliam Approach to BV Quantization}

In this section we describe a specific mathematical approach to QFT based on the work of Batalin--Vilkovisky, Wilson, Kadanoff, and others as formulated by Costello and Gwilliam\cite{CG1}.  Given such a quantum BV theory, we can extract its algebra of observables. If the theory is a \emph{cotangent theory} it determines a (projective) volume form. And finally, if the theory is translation invariant, we identify an associated $\beta$-function via RG flow.

\subsection{BV theory}

The recent text of Mnev\cite{MnevBook} provides significant motivation, context, and history for the BV formalism. We content ourselves with a streamlined presentation.

\subsubsection{BV algebras}

Recall that a {\it BV algebra} is a pair $(\cA, \Delta)$ where
\begin{itemize}
\item $\cA$ is a $\ZZ$-graded commutative associative unital algebra. 
\item $\Delta: \cA \to \cA$ is a second-order operator of degree $1$ such that $\Delta^2=0$. 
\end{itemize}
That the BV operator, $\Delta$, is ``second-order" means the following: define the \emph{BV bracket} $\{-,-\}_\Delta$ as the measuring of the failure of $\Delta$ being a derivation
\begin{equation}
    \{a,b\}_\Delta:=\Delta(ab)-(\Delta a)b- (-1)^{\lvert a \rvert}a \Delta b. 
\end{equation}
In this section we will suppress $\Delta$ from the notation, simply writing $\{-,-\}$. Then $\{-,-\}: \cA\otimes \cA\to \cA$ defines a Poisson bracket of degree $1$ satisfying
\begin{itemize}
\item $\{a,b\}=(-1)^{\lvert a \rvert \lvert b \rvert}\{b,a\}$;
\item $\{a, bc\}=\{a,b\}c+(-1)^{(\lvert a \rvert+1)\lvert b \rvert}b\{a,c\}$;
\item $\Delta\{a,b\}=-\{\Delta a, b\}-(-1)^{\lvert a \rvert}\{a, \Delta b\}$. 
\end{itemize}

A {\it differential BV algebra} is a triple $(\cA, Q, \Delta)$ where
\begin{itemize}
\item $(\cA, \Delta)$ is a BV algebra; and
\item $Q: \cA\to \cA$ is a derivation of degree $1$ such that $Q^2=0$ and $[Q, \Delta]=0$. 
\end{itemize}

Let $(\cA, Q, \Delta)$ be a differential BV algebra. A degree $0$ element $I_0\in \cA$ is said to satisfy the \emph{classical master equation} (CME) if 
\begin{equation}
QI_0+\frac{1}{2} \{I_0,I_0\}=0.
\end{equation} 
Given a solution of the CME,  the {\it complex of classical observables}, $Obs^{cl}$ is given by
\begin{equation}
Obs^{cl} \overset{\text{def}}{=}( \cA, Q+\{I_0,-\}).
\end{equation}

A degree $0$ element $I\in \cA[[\hbar]]$ is said to satisfy the \emph{quantum master equation} (QME) if 
\begin{equation}
QI+\hbar \Delta I+\frac{1}{2}\{I,I\}=0.
\end{equation}
For $(\cA, Q, \Delta)$  a differential BV algebra and $I \in \cA[[\hbar]]$  a solution of the QME, the {\it complex of quantum observables}, $Obs^{q}$ is given by
\begin{equation}
Obs^{q} \overset{\text{def}}{=}( \cA[[\hbar]], Q+\hbar \Delta + \{I,-\}).
\end{equation}

Note that $Obs^{cl}$ has a degree 1 Poisson bracket, so following \cite{CG2} we call it a $P_0$ algebra.  Similarly, in {\it ibid.} the structure on $Obs^q$ is called a BD algebra.

\subsubsection{Perturbative BV quantization}\label{app:BV2}

The data of a classical field theory over a manifold $M$ consists of a graded vector bundle $E$ (possibly of infinite rank) equipped with a -1 symplectic pairing and a local functional $S \in \cO_{loc} (\cE)$  expressed as $S(e) = \langle e, Q(e) \rangle + I_0 (e)$, where $Q$ is a square zero differential operator of cohomological degree 1, such that
\begin{itemize}
\item $S$ satisfies the CME, i.e., $\{S, S\} = 0$;
\item $I_0$ is at least cubic; and
\item $(\cE, Q)$ is an elliptic complex.
\end{itemize}

A classical field theory $(\cE, S)$ over $M$ is a {\it cotangent theory} if we can write the field content as
\begin{equation}
\cE = \Gamma \left (M ; E[1] \oplus \left (E^\vee \otimes \mathrm{Dens}(M) [-2] \right ) \right ).
\end{equation}
We further require that the action $S$ vanishes on tensors where there are at least two sections from the second summand $E^\vee \otimes \mathrm{Dens} (M)$. Cotangent theories are particularly nice: for if they admit a quantization, by a straightforward symmetry argument \cite{GGCS},  they admit a one-loop quantization.

Quantization of a field theory $(\cE, S)$ over $M$ consists of two stages:
\begin{itemize}
\item[(i)] Build a BV algebra from the data of the pairing on the bundle $E$; and
\item[(ii)] Promote the classical action $S$ to a solution of the QME in this BV algebra.
\end{itemize}

The first difficulty is that the Poisson kernel $K$ dual to the symplectic pairing is nearly always singular, so the naive definition of the BV operator $\Delta_k = \partial_k$ is ill-defined. In \cite{Cos1}, Costello uses homotopical ideas (built on the heat kernel) to build a family of well defined (smooth) BV operators $\Delta_L$ for $0<L<\infty$. Consequently, there is a family of differential BV algebras $\left \{ (\sO(E), Q , \Delta_L)\right \}_{L>0}$. Costello also describes {\it homotopy renormalization group flow} (HRG) to relate solutions of the QME between algebras in this family. As we describe in the next section,  HRG is expressed in terms of a propagator built from the differential operator $Q$ and a {\it gauge fixing operator} $Q^\dagger$; indeed, any parametrix for the generalized Laplacian $[Q, Q^\dagger]$ can be used as a propagator. 

\begin{definition}
Let $(\cE, S)$ be a classical field theory over $M$.  A {\it perturbative quantization} is a family of solutions to the QME, $\{I[L]\}_{L>0}$, linked by the HRG, such that 
\begin{equation}
\lim_{L \to 0} I[L] \equiv I_0 \quad \quad (\text{modulo } \hbar).
\end{equation}
\end{definition}

Flow via the HRG induces a chain homotopy between quantum observables  as we vary within the family of BV algebras $\{(\sO(E), Q, \Delta_L)\}_{L>0}$.  (This is one explanation for its naming convention.) Thus, we will suppress the dependence on $L$ and abusively refer to these chain homotopic complexes as the {\it global quantum observables} of our field theory.

\subsubsection{Homotopy renormalization group flow}\label{app:hrg}

The homotopy renormalization group flow equation can be described in terms of Feynman graphs. Note that our description is for an arbitrary functional on a space of fields $\cE$. Further, we will work relative to an arbitrary dg algebra $\cA$ equipped with a nilpotent ideal $\cI$.

 Let $\cO^+(\cE)\subset \cO(\cE)[[\hbar]]$
be the subspace consisting of those
functionals  which are at least cubic modulo $\hbar$ and the nilpotent ideal $\mathcal{I}$ in the base ring
$\mathcal{A}$.  Let $F\in \cO^+(\cE)$ be a functional,  which can be expanded as
\begin{equation}
F=\sum_{g,k\geq 0}\hbar^g F_{g}^{(k)}, \quad F_{g}^{(k)}\in \cO^{(k)}(\cE).
\end{equation}
We view each $F_{g}^{(k)}$ as an
$S_k$-invariant linear map $F_{g}^{(k)}: \mathcal{E}^{\otimes k}\rightarrow\mathcal{A}.$
With the propagator $P_{\epsilon \to L}$, we define the {\it (Feynman) graph weights}
\begin{equation}
W_\cG(P_{\epsilon \to L},F)\in \cO^+(\cE)
\end{equation}
for any connected graph $\cG$. We label each vertex $v$ in $\cG$ of genus $g(v)$ and valency $k$ by
$F^{(k)}_{g(v)}$. This defines an assignment $F(v):\mathcal{E}^{\otimes H(v)}\rightarrow \cA,$
where $H(v)$ is the set of half-edges of $\cG$ which are incident to $v$.
Next, we label each internal edge $e$ by the propagator 
\begin{equation}
P_e=P_{\epsilon \to L}\in\mathcal{E}^{\otimes H(e)},
\end{equation}
where $H(e)\subset H(\cG)$ is the two-element set consisting of the half-edges forming $e$. We can then contract
\begin{equation}
\otimes_{v\in V(\cG)}F(v): \mathcal{E}^{H(\cG)}\rightarrow \cA
\quad
\text{ with } 
\quad
\otimes_{e\in E(\cG)} P_e\in\mathcal{E}^{H(\cG)\setminus T(\cG)}
\end{equation}
to yield a linear map
\begin{equation}
W_\cG(P_{\epsilon \to L},F) : \mathcal{E}^{\otimes T(\cG)}\rightarrow \cA.
\end{equation}

\begin{definition}
We define the homotopy RG flow operator with respect to the propagator $P_{\epsilon \to L}$ 
\begin{equation}
   W(P_{\epsilon \to L}, -): \cO^+(\cE)\to \cO^+(\cE), 
\end{equation}
by
\begin{equation}\label{RG-flow}
W(P_{\epsilon \to L}, F):=\sum_{\cG}\frac{\hbar^{g(\cG)}}{\lvert \text{Aut}(\cG)\rvert}W_\cG(P_{\epsilon \to L}, F)
\end{equation}
where the sum is over all connected graphs.
\end{definition}


\begin{definition} A family of functionals $F[L] \in \cO^+(\cE)$ parametrized by $L>0$ is said to satisfy the homotopy renormalization group flow equation (hRGE) if for each $0 < \epsilon < L$, $F[L]=W(P_{\epsilon \to L}, F[\epsilon]).$
\end{definition}

\subsection{BV theory and volume forms: integration via homology}\label{sect:pvol}

For simplicity, let $X$ be a connected, orientable, smooth manifold of dimension $n$ Every top form $\mu \in \Omega^n(X)$ then defines a linear functional
\begin{equation}
\begin{array}{cccc}
\int_\mu:& \cinf_c(X)& \to &\RR \\
 & f & \mapsto & \int_X f \mu
 \end{array},
\end{equation}
which is a natural object from several perspectives. First, from this linear functional --- the distribution associated to $\mu$ --- we can completely reconstruct the top form $\mu$. Second, if $\mu$ is a probability measure, then $\int_\mu$ is precisely the expected value map. Our goal is now to rephrase $\int_\mu$ in a way that does not explicitly depend on ordinary integration and thus to obtain a version of volume form that can be extended to $\L8$ spaces.

We can understand $\int_\mu$ in a purely homological way, as follows. We know that integration over $X$ vanishes on total derivatives $d\omega \in \Omega^n_c(X)$, by Stokes' Theorem, so we have a commutative diagram
\begin{equation}
\xymatrix{
\Omega^n_c(X) \ar[rr]^{\int_X} \ar[rd]_{[-]} & & \RR \\
 & H^n_c(X) \ar[ru]_{\cong} &
}
\end{equation}
where $[\omega]$ denotes the cohomology class of the top form $\omega$. (The cohomology group $H^n_c(X)$ is 1-dimensional by Poincar\'e duality.) In consequence, we can identify $\int_\mu$ with the composition
\begin{equation}
\xymatrix{
 \Omega^n_c(X) \ar[r]^{[-]} &  H^n_c(X)\\
\cinf_c(X) \ar[u]^{\iota_\mu} \ar[ur]_{\int_\mu}&
}
\end{equation}
where $\iota_\mu$ denotes ``multiplication by $\mu$" (or ``contraction with $\mu$"). We thus have a purely homological version of integration against $\mu$.

It is natural to extend the map ``contract with $\mu$" to the whole de Rham complex, and not just the top forms:
\begin{equation}
\xymatrix{ 
\dotsb \ar[r] & \Omega_c^{n-2}(X) \ar[r]^d & \Omega_c^{n-1}(X) \ar[r]^d & \Omega^n_c (X)\ar[r]^{\int_X} & \RR \\
\dotsb \ar[r] & PV^2_c (X)  \ar[r]^{div_\mu} \ar[u]_{\iota_{\mu}} & PV_c^1(X)  \ar[r]^{div_\mu} \ar[u]_{\iota_{\mu}} & C_c^\infty(X) \ar[u]_{\iota_{\mu}}  \ar[ur]_{\int_\mu}
},
\end{equation}
where $PV^k_c(X) := \Gamma_c(X, \Lambda^k T_X)$ denotes the compactly-supported {\em polyvector fields} and $div_\mu$ denotes ``divergence with respect to $\mu$." We require now that $\mu$ is nowhere-vanishing, so that the divergence is well-defined. This map of cochain complexes $\iota_\mu$ is then an isomorphism.

The significance of the bottom row is that it fully encodes integration against $\mu$ but the relevant data of $\mu$ is contained in the differential $div_\mu$. Further, the bottom row often makes sense for infinite dimensional objects where the notion of ``top form" is ambiguous at best.

The complex $(PV(X), div_\mu)$ is a fundamental example of a BV algebra, the associated bracket $\{-,-\}$ is the Schouten bracket.  Further, note that we have an equivalence
\begin{equation}
 \sO (T^\ast [-1]X) \cong PV(X).
\end{equation}
Hence, when we quantize a cotangent theory we will extract a volume form (up to a scalar).

\subsection{Observable theory}

To begin, let us introduce an algebraic object which will play a central role.

\begin{definition}
Let $M$ be a manifold. A prefactorization algebra $\cF$ on $M$ with values in cochain complexes consists of
\begin{itemize}
\item[(i)] For each open $U \subseteq M$, a complex $\cF(U)$;
\item[(ii)] For any finite collection of disjoint opens $U_1, \dotsc , U_n \subset V$, where $V \subset M$ is an open, a chain map
\begin{equation}
\cF(U_1, \dotsc , U_n |V) : \cF(U_1) \otimes \cF(U_2) \otimes \dotsb \otimes \cF(U_n) \to \cF(V).
\end{equation}
\end{itemize}
Such that,
\begin{itemize}
\item[(a)] Composition is associative, i.e., for $\{T_{ij}\}$ a collection of disjoint opens in $U_i$, the following commutes

\begin{equation}
\begin{tikzcd}
\bigotimes_i \bigotimes_j \cF (T_{ij}) \ar[rr] \ar[dr] && \bigotimes_i \cF(U_i) \ar[dl] \\ & \cF(V)
\end{tikzcd}
\end{equation}

\item[(b)] The map $\cF(U_1 , \dotsc , U_n |V)$ is $S_n$ equivariant.
\end{itemize}
\end{definition}
Condition (a) is quite natural from a geometric perspective as the example below illustrates.\\

\begin{figure}
\begin{small}
 \begin{minipage}{.2\textwidth}
 \begin{center}
\begin{tikzpicture}
\draw (0,0) circle (2cm);
\draw (-.5,.5) circle (1cm);
\draw (-1,1) circle (.2cm);
\draw (-.2,1) circle (.2cm);
\draw (-.5,.2) circle (.3cm);
\draw (.9, -.5) circle (.6cm);
\draw (.9, -.5) circle (.3cm);
\node at (.9,-.5) {$T_{21}$};
\node at (.2,-1) {$U_2$};
\node at (-.5,-1.5) {$V$};
\node at (-.50,.2) {$T_{11}$};
\node at (-1,.62) {$T_{12}$};
\node at (-.05,.62) {$T_{13}$};
\node at (-1, -.6) {$U_1$};
\end{tikzpicture}
\end{center}
    \end{minipage}%
    \begin{minipage}{0.15\textwidth}
    \mbox{}
    \end{minipage}
    \begin{minipage}{0.4\textwidth}
\begin{center}
\begin{tikzcd}
\cF(T_{11}) \otimes \cF(T_{12}) \otimes \cF(T_{13}) \otimes \cF(T_{21}) \ar[dd] \ar[dr] \\ & \cF(V) \\ \cF(U_1) \otimes \cF(U_2) \ar[ur] 
\end{tikzcd}
\end{center}    
\end{minipage}
\end{small}
\caption{The generalized associativity for the structure maps of a factorization algebra.}
\end{figure}
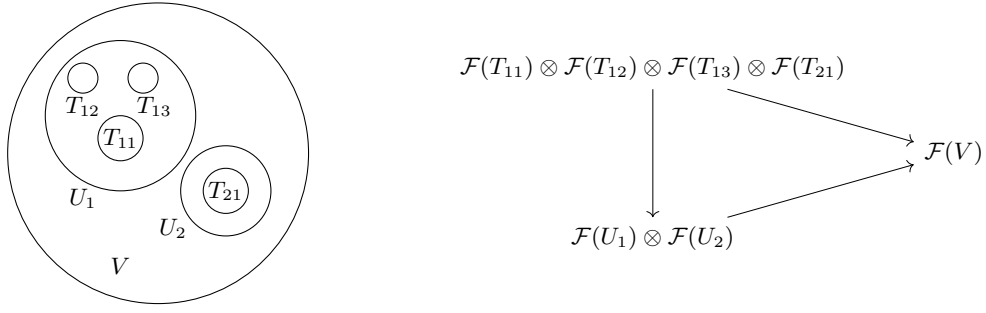

\vspace{1ex}

Note that a prefactorization algebra is similar to a cosheaf, except that it takes disjoint unions to tensor products rather than direct sums. A prefactorization algebra, $\cF$, is a {\it factorization algebra} if it is a cosheaf for the {\it Weiss topology}, i.e., $\cF$ satisfies some explicit gluing properties.

The key result in the Costello--Gwilliam\cite{CG2} paradigm is that given a quantum BV theory $(\cE,S, \Delta)$ over a manifold $M$, the quantum observables are a factorization algebra $Obs^q : \mathsf{Opens} (M) \to \mathsf{Ch}$. In fact, the classical observables, $Obs^{cl}$, also are a factorization algebra equipped with a shifted Poisson bracket (a $P_0$ structure) and $Obs^q$---or rather $Obs^q$ with its BD algebra structure---is a quantization of $Obs^{cl}$.

There is an explicit comparison between factorization algebras of observables and of nets in the pAQFT formalism of Fredenhagen and Rejzner.\cite{OwenR,FR1,FR2}

\subsubsection{Topological theories and $E_k$ algebras}

Every associative algebra $A$ defines a prefactorization algebra $\cF_A$ on $\RR$. Indeed, to any connected, nonempty,  open interval, $(a,b)$, $\cF_A (a,b) = A$. The structure maps of the factorization algebra are then induced from the algebra multiplication $m: A \otimes A \to A$.

\begin{figure}
\begin{center}
 \begin{minipage}{.2\textwidth}
 \begin{center}
\begin{tikzpicture}
\draw[line width=0.3mm, gray] (-.5,0) -- (5,0);
\draw[line width=0.3mm, gray] (-.5,2) -- (5,2);

\draw[line width=1mm] (0,0) -- (4.5,0);
\draw[line width=1mm] (1,2)--(2,2);
\draw[line width=1mm] (2.5,2)--(3.5,2);
\draw[line width=.3mm,right hook->] (1.5,1.9)--(1.5,.1);
\draw[line width=.3mm,right hook->] (3,1.9)--(3,.1);

\end{tikzpicture}
\end{center}
    \end{minipage}%
    \begin{minipage}{0.1\textwidth}
    \mbox{}
    \end{minipage}
    \begin{minipage}{0.4\textwidth}
\begin{center}
\begin{tikzcd}
A \otimes A \ar[dd,"m"] \\ \\ A
\end{tikzcd}
\end{center}    
\end{minipage}
\end{center}
\caption{The factorization algebra on $\RR$ determined by an associative algebra $A$.}
\end{figure}
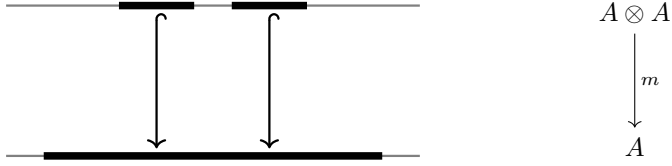

\vspace{1ex}
By construction, the factorization algebra $\cF_A$ has a special property in that the inclusion of one disc (interval) into another $D_1 \hookrightarrow D_2$ induces an equivalence $\cF_A (D_1) \simeq \cF_A (D_2)$. This behavior is common in topological field theories of arbitrary dimension. In fact, let $D \subseteq M$ be a disc in the $n$-manifold $M$ and $(\cE, \cS, \Delta)$ a BV theory over $M$ which is topological, then $Obs^q (D)$ has the structure of an $E_n$ algebra.\cite{Lurie} In the case $n=1$, an $E_1$ algebra is a homotopy associative algebra (or $A_\infty$ algebra). An example of an $E_n$ algebra for general $n$ is the algebra of cochains on an $n$-fold loop space.

\subsubsection{Holomorphic theories and vertex algebras}

The relationship between the factorization algebra of observables for holomorphic BV theories on $\CC$ and vertex algebras is spelled out in detail in Chapter 5 of Costello--Gwilliam\cite{CG1}.  The motivating picture is pretty standard, the interesting point is that the quite general notion of a factorization algebra specializes to that of a vertex algebra in this case. The underlying vector space of the vertex algebra is $\bigoplus_k Obs^q (D(0,1))_k$, the weight space decomposition of observables on the disk with respect to the $U(1)$-action.  The structure maps, e.g., OPEs, are then determined by the structure maps of the factorization algebra, i.e., for $z_1 , z_2, \dotsc , z_n \in D(0,1)$ and $0<\epsilon \ll1$ there is a map (holomorphic in the $z_i$'s)
\begin{equation}
m_{z_1 , \dotsc z_n} : Obs^q (D(z_1 , \epsilon)) \otimes \dotsb \otimes Obs^q (D(z_n, \epsilon)) \to Obs^q (D(0,1)).
\end{equation}

Williams has made this construction explicit in the case of the Virasoro algebra and describes the Segal--Sugawara map for the free $\beta \gamma$-system.\cite{Bri}

\subsection{RG flow}

Consider a translation invariant BV theory $(\cE,Q, I)$ on $\RR^n$.
The group $\RR_{>0}$ acts on the fields via the action induced by rescaling $\RR^n$. 
Further, $\RR_{>0}$ acts on the space of functionals, $\cO(\cE)$, and preserves the subspace of local functionals $\cO_{loc} (\cE)$; denote this action by $\rho_\lambda$. A classical field theory is {\it scale-invariant} if $\rho_\lambda (I) = I$ for all $\lambda \in \RR_{>0}$.

Following Ref. \cite{Cos1}, extend the action of $\RR_{>0}$ to QFTs in the BV formalism.
This action is called the {\it (local) renormalization group flow}, or simply RG flow. 
Given a quantization, so functionals $\{I[L]\}_{L>0}$ that satisfy homotopy RG flow (the renormalization group equation) and  the quantum master equation.  
Define a rescaled effective family $\{I_\lambda [L]\}$ by
\begin{equation}
I_\lambda [L] := \rho_\lambda \cdot I[\lambda^{-k} L].
\end{equation}
Elliot, Williams, and Yoo\cite{EWY} show that $\{I[L]\}$ satisfies homotopy RG flow and the QME if and only if the rescaled family $\{I_\lambda [L]\}$ does.

\begin{definition}
Let $\{I[L]\}$ be an effective quantization of the BV theory $(\cE,Q,I)$ on $\RR^n$. For $L \in \RR_{>0}$, define the {\it scale $L$ $\beta$-functional} to be the functional
\begin{equation}
\sO_\beta[L] := \lim_{\lambda \to 1} \lambda \frac{d}{d \lambda} I_\lambda [L].
\end{equation}
\end{definition}

We can expand the $\beta$-functional in powers of $\hbar$:
\begin{equation}
\sO_\beta [L] = \sO^{(0)}_\beta [L] + \hbar \sO^{(1)}_\beta [L] + O(\hbar^2).
\end{equation}
The superscript indicates the loop depth, e.g., $\sO^{(1)}_\beta [L]$ is the {\it one-loop $\beta$-functional}. For a theory where the classical theory is scale invariant, the zero-loop $\beta$-functional, $\sO^{(0)}_\beta [L]$, is identically zero.
From \cite{EWY},
\begin{equation}
\sO^{(1)}_\beta := \lim_{L \to 0} \sO^{(1)}_\beta [L]
\end{equation}
exists and determines a closed element in $\cO_{loc} (\cE)$.

A Warning:
The higher loop $\beta$-functionals, $\sO^{(k)}_\beta = \lim_{L \to 0} \sO^{(k)}_\beta [L]$, are not necessarily well-defined. 
Even if one na\"{i}vely defines $\sO^{(k)}_\beta$ to be the $\log \epsilon$ divergence at $k$-loops, these functionals are not closed with respect to the BRST differential $Q + \{I,-\}$. 
One can prove that if $\sO^{(i)}_\beta \equiv 0$ for $i < k$, then $\sO^{(k)}_\beta [L]$ satisfies homotopy RG flow and is BRST closed; in particular, the $k$-loop $\beta$-functional, $\sO^{(k)}_\beta$, exists.

Finally, we define the $\beta$-function to be the cohomology class of the $\beta$-functional:
\begin{equation}
\beta[L] := [ \sO_\beta [L]]\in H^0 (\cO(\cE) \otimes C^\infty((\epsilon, L))).
\end{equation}
Typically, the $\beta$-function cannot be decomposed by $\hbar$ degrees; the complex of functionals is only filtered, not graded.
However, at one-loop,
\begin{equation}
\beta^{(1)} := \left [\sO^{(1)}_\beta \right ] \in H^0 (\cO_{loc}(\cE))
\end{equation}
is well-defined.

In order to derive an explicit formula for $\beta^{(1)}$, we should fix {\it bare values} (a basis) for the space $H^0 (\cO_{loc} (\cE))$. Next we would like to distinguish a finite dimensional subspace of $H^0 (\cO_{loc}(\cE))$ and realize $\beta^{(1)}$ as a function of a $\RR^n$ valued parameter ${\bf c}$. Ideally, we would like to choose this distinguished  subspace to be the one spanned by the classical interaction $I$ under homotopy $RG$ flow.  However,  homotopy RG flow may not actually preserve this subspace and can introduce {\it dynamic coupling constants}. We can however restrict to the subspace spanned by functionals which are $\log \epsilon$ divergent in the $\epsilon \to 0$ limit to obtain a well-defined function of just a few coupling constants.


\section{Examples}\label{sect:examples}

In this section we will let $N$ denote a closed oriented $n$-manifold. All the examples of BV theories we discuss here are of BF or Chern--Simons type (or a deformation thereof as in the case of Yang--Mills), i.e., for $\fg$ a Lie algebra (or more generally $\L8$ algebra), the field content is given by
\begin{equation}
\cE_{BF} = \Omega^\ast_N \otimes \fg[1] \oplus \Omega^\ast_N \otimes \fg^\vee [n-2], \quad \text{ or} \quad
\cE_{CS} = \Omega^\ast_N \otimes \fg[1].
\end{equation}
For a Chern--Simons theory, it is necessary for $\fg$ to have a symmetric non-degenerate pairing. In BF theory, the pairing is induced by the Kronecker pairing between $\fg$ and $\fg^\vee$, so there is no such requirement. The action functional $S$ is the standard (perturbative) BF or Chern--Simons functional with kinetic term given by the de Rham differential. In the case of an $\L8$ algebra there are higher terms determined by the higher brackets as well as an extra kinetic term, so $Q = d_{N} \otimes 1 + 1 \otimes \ell_1$. In the curved case, i.e., $\ell_0 \neq 0$, there are some additional subtleties.\cite{LiLi}

The overriding idea is that the use of $\L8$ models allows us to  present an effective theory for the low energy sector of many $\sigma$-models as BF/Chern--Simons type theories: one uses an $\L8$ model for the appropriate mapping stack to interpret the AKSZ/PTVV action as a perturbative gauge theory. Given a target manifold, $X$, the resulting theory is actually a family of theories parametrized by the dg manifold $(X, \Omega^\ast_X)$.\cite{Cos1} This family is local in $X$, so as a consequence, the observables form a sheaf of factorization algebras over $X$.

\subsection{BF and Chern-Simons type theories}  

We will now highlight a few examples of the $\sigma$-model $\L8$ BF/Chern--Simons paradigm. There are many other interesting examples of this program: B-model\cite{LiLi}, Rozansky--Witten\cite{Qin}, BCOV\cite{KevinSi}, aspects of super gravity\cite{KevinSi2}, etc.

\subsubsection{Topological mechanics: $\hat{A}(X)$ and Fedosov quantization}

Consider topological mechanics on a cotangent bundle $T^\ast X$, so maps $\RR \to T^\ast X$ or $S^1 \to T^\ast X$. The corresponding BF theory has field content
\begin{equation}
\cE = \Omega^\ast_{S^1} \otimes \fg_X[1] \oplus \Omega^\ast_{S^1} \otimes \fg_X^\vee [-1],
\end{equation}
where $\fg_X$ is This theory is a cotangent theory, so it admits a one-loop quantization.\cite{GGCS} Moreover, observables are naturally functions on the derived loop space $\widehat{\cL_{dR} X}$, so we would expect to obtain a volume form on $\widehat{\cL_{dR} X}$ from this quantization. Indeed this is the case, and this volume form is given by $\hat{A} (X)$ which is a characteristic class of $X$ that is ubiquitous in index theory.\cite{Grady}

Now consider mechanics not on a cotangent bundle, but on a generic closed symplectic manifold $(M, \omega)$. Our space of fields is now simply $\cE = \Omega^\ast_{S^1} \otimes \fg_X$ where the pairing is induced by $\omega$. The action functional is the ($\L8$) Chern--Simons functional with kinetic term $d_{S^1} + \nabla$ for $\nabla$ a symplectic connection. This theory is not a cotangent theory and there are quantum corrections to all loop orders. However, these quantum corrections are iterative and it can be proven that solutions to the QME are in correspondence to solutions of the Fedosov equation for abelian connections.\cite{GLL} The relationship to Fedosov's work on deformation quantization\cite{Fed1,Fed2} extends to the level of observables, as the quantum observables are a deformation quantization of $\cinf (M)$ with associative product the Fedosov $\star$-product. Moreover, the geometry of the source $S^1$ leads to a trace map and recovers the {\it Algebraic Index Theorem} of Fedosov and Nest--Tsygan\cite{NT95}
\begin{equation}
\Tr (1) = \int_M e^{-\omega_\hbar/\hbar} \hat{A} (M).
\end{equation}

\subsubsection{Friedan's Theorem: Ricci flow}

Consider the two-dimensional $\sigma$-model with source a Riemann surface $\Sigma$ and target a Riemannian manifold $X$ with metric $h$. 
A field is a smooth map $\varphi : \Sigma \to X$ and the action functional is given by $S(\varphi) = \int_{\Sigma} h(\partial \varphi, \Bar{\partial} \varphi),$
where $\partial, \Bar{\partial}$ are the holomorphic and anti-holomorphic pieces of the de Rham differential on $\Sigma$. 
The map $\varphi$ satisfies the classical equations of motion if and only if it is harmonic. 
Classically the theory is conformal and when we work locally on $\Sigma$ the theory is scale invariant. 
The failure of the classical theory to be scale invariant at the quantum level is measured by the {\em $\beta$-function} of the quantum field theory. 

In 1985,  Dan Friedan \cite{Fried} gave a physical argument relating the one-loop $\beta$-function to the Ricci curvature of $X$:
\begin{equation}
\beta^{(1)} (h) = -\frac{1}{12 \pi} \mathrm{Ric} (h).
\end{equation}
With Williams\cite{GW}, we give a mathematical proof of this result in the Costello--Gwilliam formalism by using an $\L8$ model for $(X,h)$.
We do not study the full theory, but rather only the theory in perturbation around the space of constant maps in the space of all harmonic maps. Nonetheless, our low energy effective quantization still encodes much of the topology and geometry of the target manifold.

\subsubsection{Curved $\beta \gamma$: CDOs and the Witten genus} 

The curved $\beta \gamma$ system is a sigma model whose classical solutions are holomorphic maps, $\Sigma \to Y$, from a Riemann surface to a complex manifold. For $\Sigma$ an elliptic curve, Costello\cite{CosWG2} presents the low energy sector of this theory using the $\L8$ space $(Y, \fg_{Y_{\overline{\partial}}})$. The corresponding BV theory is a type of {\it holomorphic Chern--Simons} with field content $\cE = \Omega^{0,\ast}_\Sigma \otimes \fg_{Y_{\overline{\partial}}}[1] \oplus \Omega^{1,\ast}_\Sigma \otimes \fg_{Y_{\overline{\partial}}}[-1]$. This theory is a cotangent theory and admits a one-loop quantization upon trivializing $\mathrm{ch}_2 (Y)$. Moreover, the corresponding volume form is the Witten class of $Y$ determined by the curve $\Sigma$, $\mathrm{Wit}(Y,E)$.  The Witten class/genus arises as the index of ``the dirac operator" on the loop space $LY$.\cite{WitLoop,Taubes} The connection is as follows: by replacing the curve $\Sigma$ by its cohomology, Costello transforms the elliptic curve into a ``holomorphic circle."

The Witten genus is also the Euler characteristic of a sheaf of vertex algebras on a complex manifold $Y$. This sheaf (its actually a gerbe before trivialization) is the sheaf of {\it chiral differential operators} (CDOs).\cite{BD,MSV99,Cheung} Moreover, the connection between CDOs and the $\beta \gamma$ system has been known for some time.\cite{Wit07,Nekrasov} Therefore, one could hope that the quantum observables of holomorphic Chern-Simons can be identified with CDOs; Gorbounov, Gwilliam, and Williams\cite{GGW} do precisely this. The technique employed by these authors is as significant as the main result. Indeed, they use formal geometry/Gelfand--Kazhdan descent to obtain the curved $\beta \gamma$ system from a formal $\beta \gamma$ system. Such a technique has appeared before, but not at the level of factorization algebras/quantum observables and this method applies readily to any of the examples we present in the present article.

Finally, the Witten genus appears in elliptic cohomology and its equivariant flavors.  While this theory now has firm mathematical foundations,\cite{Goerss} there still is not a useable cocycle model and there is hope that methods like those just mentioned could be adapted to provide such a model. There has been some recent progress in this direction by Berwick-Evans and Tripathy.\cite{DanArnav}

\subsection{Yang--Mills theories}

Let us recall the first order formulation of (pure) Yang--Mills on $\RR^4$ following \cite{Cos1}.  Let $\cA$ denote the following dga

\begin{equation}
\begin{tikzcd}
\Omega^0 (\RR^4) \ar[r,"d"] & \Omega^1 (\RR^4) \ar[r,"d"] & \Omega^2_+ (\RR^4)\\
& \Omega^2_+ (\RR^4) \ar[ur,"2c \mathrm{Id}"] \ar[r,"d"] & \Omega^3 (\RR^4) \ar[r,"d"] & \Omega^4 (\RR^4)
\end{tikzcd}
\end{equation}
where the algebra structure is given by viewing the bottom row as a module for the top row. Let $\fg$ be a Lie algebra with invariant pairing. The action functional, $S_{FO}$, is simply given by the Chern--Simons functional on the space of fields $\cA \otimes \fg[1]$. 

Note that if $c=0$, then Yang--Mills theory can be obtained via AKSZ from $\mathrm{Map} (\RR^4, T^\ast[1] B \fg)$. Either way, there is a clear generalization to the case where $\fg$ is an $\L8$ algebra (with an invariant pairing) or more generally an $\L8$ space $(X, \fg)$. In the latter case we will obtain a family of perturbative Yang--Mills theories parametrized by $X$. These theories can be defined similarly in any even dimension. In light of our previous discussion of Lie algebroids, this provides a BV interpretation of the algebroid Yang--Mills theories of Strobl and collaborators.\cite{Str04,MayerStrobl,KotovStrobl}

\section{Boundary Conditions/Theories} 

There are several approaches to incorporating boundary and boundary conditions into the BV formalism. The program of Cattaneo, Mnev, and Reshetikhin\cite{CMR1,CMR2,CMR3,CMR4} is a BV--BFV formalism which is Lagrangian in the bulk and Hamiltonian on the boundary; a particularly illustrative example is that of {\it split Chern--Simons}.\cite{CMW} This program has been quite  successful, though for narrative consistency we will instead describe an extension of the Costello--Gwilliam approach which incorporates boundary conditions.\cite{AR,BY, GRW} 

\subsection{Derived lagrangians}

Following Ref \cite{CalaqueLectures} we introduce Lagrangian structures in the derived setting. Again we will see that being Lagrangian is structure, not a property. 

Let $(X, \omega)$ be a $n$-shifted symplectic space/stack and $f: L \to X$ a map. An {\it isotropic structure} on $f$ is a homotopy $\gamma$ between $f^\ast \omega$ and $0$ in $\Omega^{2,cl} (L)$. Explicitly, $\gamma$ is given by a sequence of forms $(\gamma_0, \gamma_1, \dotsc)$ where
\begin{itemize}
\item For each $i$, $\gamma_i \in \Omega^{2+i} (L)$ and is of degree $1+n$; and
\item $\left ( d_{dR} + \partial \right ) \gamma = f^\ast \omega$, i.e., $\partial \gamma_0 = f^\ast \omega_0$, and $\partial \gamma_{i+1} + d_{dR} \gamma_i = f^\ast \omega_{i+1}$ for all $i\ge0$.
\end{itemize}
So $\gamma$ corresponds to a choice of a coboundary for $f^\ast \omega$ in $\Omega^{2,cl} (L)$. Define the {\it normal complex} of $f$, $\NN_f$, by
\begin{equation}
\NN_f := \mathrm{cone} \left ( \TT_L \to f^\ast \TT_X \right ).
\end{equation}
An isotropic structure structure determines a map $\NN_f \to \LL_L [n]$.

\begin{definition}
An isotropic structure on the map $f: L \to X$ is Lagrangian if the induced map $\NN_f \to \LL_L [n]$ is an equivalence.
\end{definition}

It is an interesting exercise to verify directly that a Lagrangian $f: L \to \ast_{(n)}$ to the point with its unique $n$-shifted symplectic structure is a $(n-1)$-shifted symplectic structure on $L$.

\subsubsection{Lagrangian intersections}

Let $X$ be $n$-shifted symplectic and $f_1: L_1 \to X$, $f_2 :L_2 \to X$ a pair of Lagrangians. Then we can consider the (derived) intersection/pullback $L_1 \times_X L_2$; it is a fundamental result that this intersection is itself $(n-1)$-shifted symplectic.

As a special case of the proceeding we consider the (derived) critical locus. Let $X$ be a manifold and $S: X \to \mathbb{K}$  a function. The graph of $dS$ and the zero section $X \hookrightarrow T^\ast X$ are Lagrangians and their intersection is the {\it derived critical locus} of $S$, $\mathrm{dCrit}(S)$. It follows from the preceding paragraph that $\mathrm{dCrit}(S)$ is (-1)-symplectic. As perturbative BV studies $\mathrm{dCrit}(S)$, we see again the connection to (-1)-shifted symplectic geometry. Complimentarily, one could motivate derived symplectic geometry by the desire to study $\mathrm{dCrit}(S)$ for functions defined on domains that are not necessarily smooth manifolds or schemes.

\subsubsection{Twisted Dirac structures as Lagrangians}

Roytenberg studied $n$-shifted symplectic structures on dg manifolds and in his classification they corresponded to symplectic, Poisson, and Courant algebroid structures respectively for $n=0,1,2$.\cite{Roy2} In the more general framework for derived geometry presented above, Pym and Safronov classified $n$-shifted symplectic algebroids over a manifold or scheme.\cite{PS} The examples of Roytenberg arise as isotropic structures in the more general setting. In the case $n=2$, symplectic algebroids correspond to {\it twisted} Courant algebroids.  

A prototypical example of a (higher) Courant algebroid is that of an exact one twisted by a differential form over a manifold $X$.
There is a {\it standard} $k$-Courant bracket on $T_X \oplus \Lambda^{k} T^\ast_X$ given by
\begin{equation}
[[A + \lambda , B + \xi]]:=[A,B]+ \cL_A \xi - \cL_B \lambda +\frac{1}{2} d \left ( \iota_B \lambda - \iota_A \xi \right ).
\end{equation}
This bracket can be twisted by a $(k+2)$-flux $G$:
\begin{equation}
[[A + \lambda , B + \xi]]:=[A,B]+ \cL_A \xi - \cL_B \lambda +\frac{1}{2} d \left ( \iota_B \lambda - \iota_A \xi \right ) + \iota_A \iota_B G.
\end{equation}
These standard exact algebroids determine $n$-symplectic spaces where $n=k+2$.

Let $\sE$ be (a possibly twisted) Courant algebroid over a manifold or variety $X$.  As $\sE$ is 2-shifted symplectic, we could ask to classify the Lagrangians of $\sE$. Pym and Safronov prove that Dirac structures supported on a closed submanifold $Y \subset X$ of an untwisted $\sE$ classify isotropic structures, so they need not be non-degenerate.  However, in the twisted case they do obtain an equivalence of groupoids between {\it twisted Dirac pairs} and the ($\infty$) groupoid of 2-shifted symplectic algebroids over $X$ equipped with a Lagrangian supported on $Y$.

\subsection{BV with boundary}

We now give an impressionistic description of BV on a manifold with boundary. (All necessary modifications are spelled out in \cite{AR} and \cite{GRW}.) Let $M$ be a manifold with boundary, $\partial M$, and $i : \partial M \hookrightarrow M$ the inclusion map. Let $(\cE, S)$ be a classical field theory over $M$ and $\cE_\partial$ the restriction of fields to the boundary. Note that $\cE_\partial$ is a sheaf of complexes on $\partial M$ equipped with a 0-shifted symplectic form. Let $\cL \subset \cE_\partial$ be a Lagrangian specifying the boundary condition. The {\it space of $\cL$-conditioned fields} $\cE_\cL$ is the obtained via pullback of sheaves of complexes over $M$
\begin{equation}
\begin{tikzcd}
    \cE_\cL \arrow[r] \arrow[d]
    \arrow[dr, phantom, "\ulcorner", very near start]
    & \cE \arrow[d] \\
    i_\ast \cL \arrow[hookrightarrow]{r}
    & i_\ast \cE_\partial
\end{tikzcd}
\end{equation}

The conditioned fields $\cE_\cL$ has a (-1)-symplectic structure. This can be verified directly as its a subsheaf of $\cE$. Alternatively, this follows from the previous section as $\cE_\cL$ is the derived intersection of two lagrangians in a (0)-shifted symplectic space (some care is needed to make this homotopically coherent\cite{AR}). While it is not necessarily the case that the sheaf $\cE_\cL$ is sections of a vector bundle over $M$, much of the Costello--Gwilliam approach still applies to this {\it bulk-boundary system}.

For bulk-boundary systems, the observables (both classical and quantum) form a {stratified factorization algebra}. In particular, their is an action of the bulk observables on the boundary observables. This stratified factorization algebra for the linear Poisson sigma model recovers the Swiss cheese algebras which appear in formality/deformation quantization.  Similarly, Koszul duality appears when one considers a pair of transverse lagrangians.\cite{GRW,CFFR,Shoiket}

\subsection{CS/WZW correspondence}

Let us describe a perturbative description of the Chern--Simons/Wess--Zumino--Witten correspondence in the abelian case.\cite{GRW} Let $V$ be a complex vector space equipped with a symmetric bilinear pairing $\kappa$ which is non-degenerate. Let $M=\Sigma \times \RR_{\ge 0}$ for $\Sigma$ a Riemann surface. As before, perturbative Chern--Simons has fields $\Omega^\ast_M \otimes V[1]$, though here it is useful to decompose fields as
\begin{equation}
\cE = \Omega^{0,\ast}_\Sigma \otimes \Omega^\ast_{\RR \ge 0} \otimes V[1] \oplus  \Omega^{1,\ast}_\Sigma \otimes \Omega^\ast_{\RR \ge 0} \otimes V.
\end{equation}
The differential is given by $Q= \partial + \overline{\partial} + d_{dR}$ and $\kappa$ induces a local pairing. Consider the {\it chiral WZW boundary condition} determined by
\begin{equation}
\Omega^{1,\ast}_\Sigma \otimes V \subset \Omega^\ast_\Sigma \otimes V[1].
\end{equation}

Upon quantizing this (free) bulk-boundary system, one obtains a stratified factorization algebra that interprets between the chiral $U(1)$ currents on the boundary (complete with its OPE and structure of a vertex algebra) and the bulk Chern--Simons observables which are a certain twisted factorization envelope.\cite{GRW} Mnev, Schiavina, and Wernli obtain related results working in the BV--BFV formalism.\cite{MSW}

\subsection{A new proof of the Bursztyn--Dolgushev--Waldmann Theorem}

By using Hochschild homology techniques, Bursztyn, Dolgushev, and Waldmann\cite{Dol} (BDW) proved a conjecture of Chervov and Rybnikov\cite{CRconj} that for a symplectic manifold the Kontsevich $\star$-product\cite{Kontsevich} and the Fedosov $\star$-product\cite{Fed1} agree. BDW's approach is largely algebraic and it seemed desirable to connect directly to the work of Cattaneo and Felder on the Poisson $\sigma$-model.\cite{CF1,CF2} 

Indeed the Poisson $\sigma$-model for a symplectic manifold provides another proof of BDW Theorem (though the underlying geometric picture was certainly known to Dolgushev and collaborators among others, so ``another" should be taken with a grain of salt.). From \cite{GGAlgd}, we know that there corresponds an $\L8$ space encoding a Poisson manifold $(M,\Pi)$; in. the examples above this space was denoted $(M, \fg_\Pi)$. Further, $(M, \fg_\Pi)$ is 1-symplectic so serves as the target of a two dimensional $\sigma$-model.  Cui and Zhu\cite{CZ} quantize this theory in the BV formalism for a surface with boundary in the Costello--Gwilliam formalism. 

When $\Pi$ comes from a symplectic structure,  the bulk theory is acyclic and is equivalent to topological mechanics on the boundary.  Restricting to the case where the source manifold is the disc, this latter theory is the one discussed above and  \cite{GLL} shows that factorization algebra structure on boundary observables returns the Fedosov $\star$-product. The comparison result then follows from Refs. \cite{CF1,CF2} as the the boundary observables are shown to encode the Kontsevich $\star$-product.

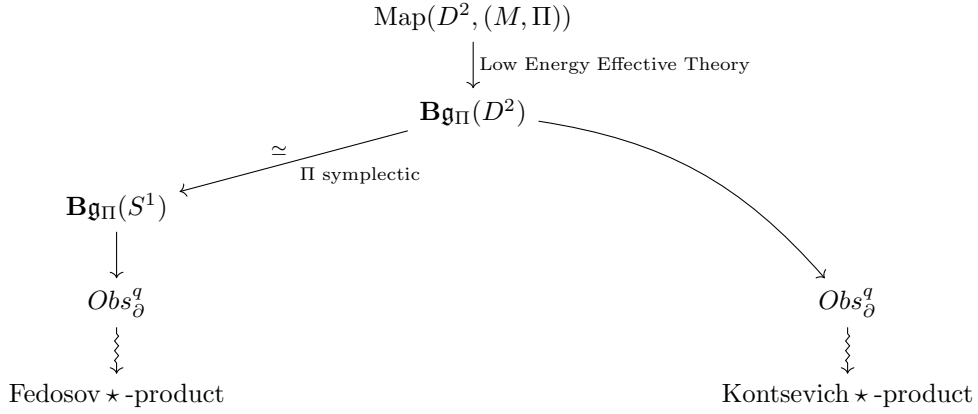
\begin{figure}
\begin{tikzcd}
&&\mathrm{Map} (D^2 , (M, \Pi)) \ar[d,"\text{Low Energy Effective Theory}"]\\
&&\bB \fg_\Pi (D^2) \ar[dll, "\Pi \text{ symplectic}","\simeq"'] \ar[rrdd,bend left=20]\\
\bB \fg_\Pi (S^1) \ar[d]\\
Obs^q_\partial \ar[d,rightsquigarrow]&&&& Obs^q_\partial \ar[d, rightsquigarrow]\\
\text{Fedosov} \star \text{-product} &&&& \text{Kontsevich} \star \text{-product}
\end{tikzcd}
\caption{A schematic illustrating the equivalence of the Fedosov $\star$-product and the Kontsevich $\star$-product.}
\end{figure}


\section{Character Varieties: A View of the AJ Conjecture}

Imprecisely, the {\it AJ Conjecture} of Garoufalidis\cite{Gar} states that the colored Jones polynomial of a knot determines its A-polynomial. Via $SL_2 (\CC)$ character varieties this conjecture is related to questions of hyperbolic volume and slopes of incompressible surfaces in the knot complement. The conjecture is known to hold for most 2-bridge knots and some selected classes of other knots.\cite{LeZhang} Given a knot manifold $N$, there is a commutative diagram

\begin{equation}\label{aj}
\begin{tikzcd}
\cS (\partial N) \ar[r] \ar[d] & \cS (N) \ar[d] \\ \CC[X(\partial N)] \ar[r] & \CC[X(N)]
\end{tikzcd}
\end{equation}
The kernel of top map between skein modules encodes the colored Jones polynomial, while the map between character rings of character varieties encodes the A-polynomial. The AJ conjecture---very roughly---is the comparison of kernels.

In this section we propose a way of studying the AJ conjecture through the lens of derived symplectic geometry.

\subsection{Character varieties and their shifted symplectic structure}

Let $N$ be a knot manifold, i.e., a connected, compact, irreducible, orientable 3-manifold whose boundary is an incompressible torus. Such knot manifolds arise as the complement of a regular neighborhood of a knot in a homology 3-sphere.
Recall that the character variety, $X(N)$, is the affine algebraic set obtained by considering characters of all representations of $\pi (N) \to SL_2 (\CC)$. Culler and Schalen\cite{CS83} developed character varieties to study essential surfaces in $N$. Most significantly, the $A$-polynomial is a knot invariant which is determined by the regular map $i^\ast : X(N) \to X(\partial N)$.\cite{CCGL,Chesebro}

Let us replace the character variety by the character stack, $\widetilde{X} (\partial N)$, where
\begin{equation}
\widetilde{X} (\partial N) := \mathrm{Map} (\pi_1 (\partial N) , \cL BSL_2 (\CC)).
\end{equation}
The character stack is shifted symplectic as $BSL_2 (\CC)$ has a 2-symplectic structure provided by the Killing form. The first, seemingly doable, task is to equip the map $i^\ast : \widetilde{X} (N) \to \widetilde{X} (\partial N)$ with a Lagrangian structure.\cite{Calaque13}

\subsection{The quantum nature of Skein modules}

Skein modules were introduced by Przytycki\cite{Prz} and Turaev\cite{Turaev} in order to formulate the (colored) Jones polynomial for knots in an arbitrary compact oriented 3-manifold $M$. The skein module of $M$ is the free $\CC[[\hbar]]$-module with basis the set of framed links in $M$, modulo all possible skein and framing relations.  We extend this definition to compact oriented surfaces by crossing with the interval.

Work of Frohman and collaborators\cite{BFKB, FrohmanGelca} demonstrates that $\cS (\partial N \times I)$ is a deformation quantization of $\CC[X(\partial N)]$. Explicitly, their work realizes the skein algebra $\cS (\partial N \times I)$ as a subalgebra of the noncommutative torus.

\subsection{The AJ conjecture}

Our proposed method of attacking the AJ conjecture is the following:
\begin{itemize}
\item[(a)] Equip the restriction map $\widetilde{X} (N) \to \widetilde{X} (\partial N)$ with a Lagrangian structure;
\item[(b)] Realize $\cS (N)$ as a quantization of this Lagrangian which is compatible with $\cS (\partial N \times I)$, e.g., as a DQ-module;
\item[(c)] Exploit a derived Darboux type lemma to make the resulting maps (and their kernels) in Diagram \ref{aj} calculable and explicit.
\end{itemize}

\section*{Acknowledgments}
We thank David Ayala, Damien Calaque, Dan Berwick-Evans, Owen Gwilliam, Charles Katerba, Qin Li, Si Li, and Brian Williams for discussion and collaboration related to work in this note.

\bibliographystyle{alpha}
\bibliography{map_stack}

\newcommand{\etalchar}[1]{$^{#1}$}
\begin{thebibliography}{CCG{\etalchar{+}}94}

\bibitem[AR]{AR}
Benjamin~I. Albert and Eugene Rabinovich.
\newblock Factorization algebras for quantum field theories on manifolds with
  boundary,.
\newblock preprint, 2020.

\bibitem[ASZK97]{AKSZ}
M.~Alexandrov, A.~Schwarz, O.~Zaboronsky, and M.~Kontsevich.
\newblock The geometry of the master equation and topological quantum field
  theory.
\newblock {\em Internat. J. Modern Phys. A}, 12(7):1405--1429, 1997.

\bibitem[BD04]{BD}
Alexander Beilinson and Vladimir Drinfeld.
\newblock {\em Chiral algebras}, volume~51 of {\em American Mathematical
  Society Colloquium Publications}.
\newblock American Mathematical Society, Providence, RI, 2004.

\bibitem[BDW12]{Dol}
H.~Bursztyn, V.~Dolgushev, and S.~Waldmann.
\newblock Morita equivalence and characteristic classes of star products.
\newblock {\em J. Reine Angew. Math.}, 662:95--163, 2012.

\bibitem[BET]{DanArnav}
Daniel Berwick-Evans and Arnav Tripathy.
\newblock A model for complex analytic equivariant elliptic cohomology from
  quantum field theory.
\newblock available at
  \href{https://arxiv.org/abs/1805.04146}{arXiv:1805.04146}.

\bibitem[BFKB99]{BFKB}
Doug Bullock, Charles Frohman, and Joanna Kania-Bartoszy\'{n}ska.
\newblock Understanding the {K}auffman bracket skein module.
\newblock {\em J. Knot Theory Ramifications}, 8(3):265--277, 1999.

\bibitem[BY]{BY}
Dylan Butson and Philsang Yoo.
\newblock Degenerate classical field theories and boundary theories.
\newblock available at
  \href{https://arxiv.org/abs/1611.00311}{arXiv:1611.00311}.

\bibitem[BZN12]{BZN}
David Ben-Zvi and David Nadler.
\newblock Loop spaces and connections.
\newblock {\em J. Topol.}, 5(2):377--430, 2012.

\bibitem[Cal]{Damien18}
Damien Calaque.
\newblock Derived stacks in symplectic geometry.
\newblock to appear in {\it New Spaces for Mathematics and Physics} eds.~M.
  Anel and G. Catren.

\bibitem[Cal14]{CalaqueLectures}
Damien Calaque.
\newblock Three lectures on derived symplectic geometry and topological field
  theories.
\newblock {\em Indag. Math. (N.S.)}, 25(5):926--947, 2014.

\bibitem[Cal15]{Calaque13}
Damien Calaque.
\newblock Lagrangian structures on mapping stacks and semi-classical {TFT}s.
\newblock In {\em Stacks and categories in geometry, topology, and algebra},
  volume 643 of {\em Contemp. Math.}, pages 1--23. Amer. Math. Soc.,
  Providence, RI, 2015.

\bibitem[CCG{\etalchar{+}}94]{CCGL}
D.~Cooper, M.~Culler, H.~Gillet, D.~D. Long, and P.~B. Shalen.
\newblock Plane curves associated to character varieties of {$3$}-manifolds.
\newblock {\em Invent. Math.}, 118(1):47--84, 1994.

\bibitem[CF00]{CF1}
Alberto~S. Cattaneo and Giovanni Felder.
\newblock A path integral approach to the {K}ontsevich quantization formula.
\newblock {\em Comm. Math. Phys.}, 212(3):591--611, 2000.

\bibitem[CF01]{CF2}
Alberto~S. Cattaneo and Giovanni Felder.
\newblock On the {AKSZ} formulation of the {P}oisson sigma model.
\newblock volume~56, pages 163--179. 2001.
\newblock EuroConf\'{e}rence Mosh\'{e} Flato 2000, Part II (Dijon).

\bibitem[CFFR11]{CFFR}
Damien Calaque, Giovanni Felder, Andrea Ferrario, and Carlo~A. Rossi.
\newblock Bimodules and branes in deformation quantization.
\newblock {\em Compos. Math.}, 147(1):105--160, 2011.

\bibitem[CG]{CG2}
Kevin Costello and Owen Gwilliam.
\newblock Factorization algebras in quantum field theory. {V}ol. 2.
\newblock to appear.

\bibitem[CG17]{CG1}
Kevin Costello and Owen Gwilliam.
\newblock {\em Factorization algebras in quantum field theory. {V}ol. 1},
  volume~31 of {\em New Mathematical Monographs}.
\newblock Cambridge University Press, Cambridge, 2017.

\bibitem[Che05]{Chesebro}
Eric Chesebro.
\newblock All roots of unity are detected by the {$A$}-polynomial.
\newblock {\em Algebr. Geom. Topol.}, 5:207--217, 2005.

\bibitem[Che12]{Cheung}
Pokman Cheung.
\newblock Chiral differential operators on supermanifolds.
\newblock {\em Math. Z.}, 272(1-2):203--237, 2012.

\bibitem[CJLS19]{Szabo}
Athanasios Chatzistavrakidis, Larisa Jonke, Dieter L\"{u}st, and Richard~J.
  Szabo.
\newblock Fluxes in exceptional field theory and threebrane sigma-models.
\newblock {\em J. High Energy Phys.}, (5):055, 33, 2019.

\bibitem[CLa]{KevinSi}
Kevin Costello and Si~Li.
\newblock Quantum bcov theory on calabi--yau manifolds and the higher genus
  b-model.
\newblock available at \href{https://arxiv.org/abs/1201.4501}{arXiv:1201.4501}.

\bibitem[CLb]{KevinSi2}
Kevin Costello and Si~Li.
\newblock Twisted supergravity and its quantization.
\newblock available at
  \href{https://arxiv.org/abs/1606.00365}{arXiv:1606.00365}.

\bibitem[CLL17]{Qin}
Kwokwai Chan, Naichung~Conan Leung, and Qin Li.
\newblock B{V} quantization of the {R}ozansky-{W}itten model.
\newblock {\em Comm. Math. Phys.}, 355(1):97--144, 2017.

\bibitem[CMR14]{CMR1}
Alberto~S. Cattaneo, Pavel Mnev, and Nicolai Reshetikhin.
\newblock Classical {BV} theories on manifolds with boundary.
\newblock {\em Comm. Math. Phys.}, 332(2):535--603, 2014.

\bibitem[CMR18a]{CMR3}
Alberto~S. Cattaneo, Pavel Mnev, and Nicolai Reshetikhin.
\newblock Perturbative quantum gauge theories on manifolds with boundary.
\newblock {\em Comm. Math. Phys.}, 357(2):631--730, 2018.

\bibitem[CMR18b]{CMR2}
Alberto~S. Cattaneo, Pavel Mnev, and Nicolai Reshetikhin.
\newblock Poisson sigma model and semiclassical quantization of integrable
  systems.
\newblock {\em Rev. Math. Phys.}, 30(6):1840004, 26, 2018.

\bibitem[CMR20]{CMR4}
Alberto~S. Cattaneo, Pavel Mnev, and Nicolai Reshetikhin.
\newblock A cellular topological field theory.
\newblock {\em Comm. Math. Phys.}, 374(2):1229--1320, 2020.

\bibitem[CMW17]{CMW}
Alberto~S. Cattaneo, Pavel Mnev, and Konstantin Wernli.
\newblock Split {C}hern-{S}imons theory in the {BV}-{BFV} formalism.
\newblock In {\em Quantization, geometry and noncommutative structures in
  mathematics and physics}, Math. Phys. Stud., pages 293--324. Springer, Cham,
  2017.

\bibitem[Cos]{CosWG2}
Kevin Costello.
\newblock A geometric construction of the {W}itten genus, {II}.
\newblock available at
  \href{http://front.math.ucdavis.edu/1112.0816}{arXiv:1112.0816}.

\bibitem[Cos11]{Cos1}
Kevin Costello.
\newblock {\em Renormalization and effective field theory}, volume 170 of {\em
  Mathematical Surveys and Monographs}.
\newblock American Mathematical Society, Providence, RI, 2011.

\bibitem[CR]{CRconj}
A.~Chervov and L.~Rybnikov.
\newblock Deformation quantization of submanifolds and reductions via
  duflo--kirillov--kontsevich map.
\newblock available at
  \href{https://arxiv.org/abs/hep-th/0409005}{hep-th/0409005}.

\bibitem[CS83]{CS83}
Marc Culler and Peter~B. Shalen.
\newblock Varieties of group representations and splittings of {$3$}-manifolds.
\newblock {\em Ann. of Math. (2)}, 117(1):109--146, 1983.

\bibitem[CS11]{CS}
Alberto~S. Cattaneo and Florian Sch{\"a}tz.
\newblock Introduction to supergeometry.
\newblock {\em Rev. Math. Phys.}, 23(6):669--690, 2011.

\bibitem[CZ]{CZ}
Xiaoyi Cui and Chenchang Zhu.
\newblock A holography theory of poisson sigma model and deformation
  quantization.
\newblock available at
  \href{https://arxiv.org/abs/2004.00774}{arXiv:2004.00774}.

\bibitem[EWY18]{EWY}
Chris Elliott, Brian Williams, and Philsang Yoo.
\newblock Asymptotic freedom in the {BV} formalism.
\newblock {\em J. Geom. Phys.}, 123:246--283, 2018.

\bibitem[Fed94]{Fed1}
Boris~V. Fedosov.
\newblock A simple geometrical construction of deformation quantization.
\newblock {\em J. Differential Geom.}, 40(2):213--238, 1994.

\bibitem[Fed96]{Fed2}
Boris Fedosov.
\newblock {\em Deformation quantization and index theory}, volume~9 of {\em
  Mathematical Topics}.
\newblock Akademie Verlag, Berlin, 1996.

\bibitem[FG00]{FrohmanGelca}
Charles Frohman and R\u{a}zvan Gelca.
\newblock Skein modules and the noncommutative torus.
\newblock {\em Trans. Amer. Math. Soc.}, 352(10):4877--4888, 2000.

\bibitem[FR12]{FR2}
Klaus Fredenhagen and Katarzyna Rejzner.
\newblock Batalin-{V}ilkovisky formalism in the functional approach to
  classical field theory.
\newblock {\em Comm. Math. Phys.}, 314(1):93--127, 2012.

\bibitem[FR13]{FR1}
Klaus Fredenhagen and Katarzyna Rejzner.
\newblock Batalin-{V}ilkovisky formalism in perturbative algebraic quantum
  field theory.
\newblock {\em Comm. Math. Phys.}, 317(3):697--725, 2013.

\bibitem[Fri85]{Fried}
Daniel~Harry Friedan.
\newblock Nonlinear models in {$2+\varepsilon$} dimensions.
\newblock {\em Ann. Physics}, 163(2):318--419, 1985.

\bibitem[Gar04]{Gar}
Stavros Garoufalidis.
\newblock On the characteristic and deformation varieties of a knot.
\newblock In {\em Proceedings of the {C}asson {F}est}, volume~7 of {\em Geom.
  Topol. Monogr.}, pages 291--309. Geom. Topol. Publ., Coventry, 2004.

\bibitem[Get09]{Getzler}
Ezra Getzler.
\newblock Lie theory for nilpotent {$L\sb \infty$}-algebras.
\newblock {\em Ann. of Math. (2)}, 170(1):271--301, 2009.

\bibitem[GG14]{GGCS}
Ryan Grady and Owen Gwilliam.
\newblock One-dimensional {C}hern--{S}imons theory and the \^{A} genus.
\newblock {\em Algebr. Geom. Topol.}, 14(4):419--497, 2014.

\bibitem[GG15]{GGLoop}
Ryan Grady and Owen Gwilliam.
\newblock {$L\sb \infty$} spaces and derived loop spaces.
\newblock {\em New York J. Math.}, 21:231--272, 2015.

\bibitem[GG20]{GGAlgd}
Ryan Grady and Owen Gwilliam.
\newblock Lie algebroids as {$L_\infty$} spaces.
\newblock {\em J. Inst. Math. Jussieu}, 19(2):487--535, 2020.

\bibitem[GGW]{GGW}
Vasily Gorbounov, Owen Gwilliam, and Brian Williams.
\newblock Chiral differential operators via quantization of the holomorphic
  $\sigma$-model.
\newblock to appear in {\it Asterique}.

\bibitem[GLL17]{GLL}
Ryan~E. Grady, Qin Li, and Si~Li.
\newblock Batalin--{V}ilkovisky quantization and the algebraic index.
\newblock {\em Adv. Math.}, 317:575--639, 2017.

\bibitem[Goe10]{Goerss}
Paul~G. Goerss.
\newblock Topological modular forms [after {H}opkins, {M}iller and {L}urie].
\newblock Number 332, pages Exp. No. 1005, viii, 221--255. 2010.
\newblock S\'{e}minaire Bourbaki. Volume 2008/2009. Expos\'{e}s 997--1011.

\bibitem[GR17]{GR}
Dennis Gaitsgory and Nick Rozenblyum.
\newblock {\em A study in derived algebraic geometry. {V}ol. {II}.
  {D}eformations, {L}ie theory and formal geometry}, volume 221 of {\em
  Mathematical Surveys and Monographs}.
\newblock American Mathematical Society, Providence, RI, 2017.

\bibitem[GR20]{OwenR}
Owen Gwilliam and Kasia Rejzner.
\newblock Relating nets and factorization algebras of observables: free field
  theories.
\newblock {\em Comm. Math. Phys.}, 373(1):107--174, 2020.

\bibitem[Gra18]{Grady}
Ryan Grady.
\newblock The {$\hat A$}-genus as a projective volume form on the derived loop
  space.
\newblock {\em Math. Phys. Anal. Geom.}, 21(2):Paper No. 13, 30, 2018.

\bibitem[GRW]{GRW}
Owen Gwilliam, Eugene Rabinovich, and Brian Williams.
\newblock Factorization algebras and abelian cs/wzw-type correspondences,.
\newblock available at
  \href{https://arxiv.org/abs/2001.07888}{arXiv:2001.07888}.

\bibitem[GW18]{GW}
Ryan Grady and Brian Williams.
\newblock Homotopy {RG} flow and the non-linear {$\sigma$}-model.
\newblock In {\em Topology and quantum theory in interaction}, volume 718 of
  {\em Contemp. Math.}, pages 187--211. Amer. Math. Soc., Providence, RI, 2018.

\bibitem[Kon03]{Kontsevich}
Maxim Kontsevich.
\newblock Deformation quantization of {P}oisson manifolds.
\newblock {\em Lett. Math. Phys.}, 66(3):157--216, 2003.

\bibitem[KS15]{KotovStrobl}
Alexei Kotov and Thomas Strobl.
\newblock Curving {Y}ang-{M}ills-{H}iggs gauge theories.
\newblock {\em Phys. Rev. D}, 92(8):085032, 5, 2015.

\bibitem[LL16]{LiLi}
Qin Li and Si~Li.
\newblock On the {B}-twisted topological sigma model and {C}alabi--{Y}au
  geometry.
\newblock {\em J. Differential Geom.}, 102(3):409--484, 2016.

\bibitem[Lur]{Lurie}
Jacob Lurie.
\newblock Higher algebra.
\newblock to appear.

\bibitem[LZ17]{LeZhang}
Thang T.~Q. L\^{e} and Xingru Zhang.
\newblock Character varieties, {$A$}-polynomials and the {AJ} conjecture.
\newblock {\em Algebr. Geom. Topol.}, 17(1):157--188, 2017.

\bibitem[Mne19]{MnevBook}
Pavel Mnev.
\newblock {\em Quantum field theory: {B}atalin-{V}ilkovisky formalism and its
  applications}, volume~72 of {\em University Lecture Series}.
\newblock American Mathematical Society, Providence, RI, 2019.

\bibitem[MS09]{MayerStrobl}
C.~Mayer and T.~Strobl.
\newblock Lie algebroid {Y}ang-{M}ills with matter fields.
\newblock {\em J. Geom. Phys.}, 59(12):1613--1623, 2009.

\bibitem[MSV99]{MSV99}
Fyodor Malikov, Vadim Schechtman, and Arkady Vaintrob.
\newblock Chiral de {R}ham complex.
\newblock {\em Comm. Math. Phys.}, 204(2):439--473, 1999.

\bibitem[MSW]{MSW}
Pavel Mnev, Michele Schiavina, and Konstantin Wernli.
\newblock Towards holography in the bv--bfv setting.
\newblock available at
  \href{https://arxiv.org/abs/1905.00952}{arXiv:1905.00952}.

\bibitem[Nek]{Nekrasov}
Nikita Nekrasov.
\newblock Lectures on curved beta-gamma systems, pure spinors, and anomalies.
\newblock available at
  \href{https://arxiv.org/abs/hep-th/0511008}{hep-th/0511008}.

\bibitem[NT95]{NT95}
Ryszard Nest and Boris Tsygan.
\newblock Algebraic index theorem.
\newblock {\em Comm. Math. Phys.}, 172(2):223--262, 1995.

\bibitem[Prz91]{Prz}
J\'{o}zef~H. Przytycki.
\newblock Skein modules of {$3$}-manifolds.
\newblock {\em Bull. Polish Acad. Sci. Math.}, 39(1-2):91--100, 1991.

\bibitem[PS]{PS}
Brent Pym and Pavel Safronov.
\newblock Shifted symplectic lie algebroids.
\newblock to appear in {\it Int. Math. Res. Not.}

\bibitem[PTVV13]{PTVV}
Tony Pantev, Bertrand To{\"e}n, Michel Vaqui{\'e}, and Gabriele Vezzosi.
\newblock Shifted symplectic structures.
\newblock {\em Publ. Math. Inst. Hautes \'Etudes Sci.}, 117:271--328, 2013.

\bibitem[PY]{PSY}
P.~Pulmann, J.~\v{S}evera and D.~Youmans.
\newblock Renormalization group flow of chern--simons boundary conditions and
  generalized ricci tensor,.
\newblock available at
  \href{https://arxiv.org/abs/2009.00509}{arXiv:2009.00509}.

\bibitem[Roy02]{Roy1}
Dmitry Roytenberg.
\newblock On the structure of graded symplectic supermanifolds and {C}ourant
  algebroids.
\newblock In {\em Quantization, {P}oisson brackets and beyond ({M}anchester,
  2001)}, volume 315 of {\em Contemp. Math.}, pages 169--185. Amer. Math. Soc.,
  Providence, RI, 2002.

\bibitem[Roy07]{Roy2}
Dmitry Roytenberg.
\newblock A{KSZ}-{BV} formalism and {C}ourant algebroid-induced topological
  field theories.
\newblock {\em Lett. Math. Phys.}, 79(2):143--159, 2007.

\bibitem[Sho10]{Shoiket}
Boris Shoikhet.
\newblock Koszul duality in deformation quantization and {T}amarkin's approach
  to {K}ontsevich formality.
\newblock {\em Adv. Math.}, 224(3):731--771, 2010.

\bibitem[Str04]{Str04}
Thomas Strobl.
\newblock Algebroid {Y}ang-{M}ills theories.
\newblock {\em Phys. Rev. Lett.}, 93(21):211601, 4, 2004.

\bibitem[Tau89]{Taubes}
Clifford~Henry Taubes.
\newblock {$S^1$} actions and elliptic genera.
\newblock {\em Comm. Math. Phys.}, 122(3):455--526, 1989.

\bibitem[Toen14]{Toen}
Bertrand To\"{e}n.
\newblock Derived algebraic geometry.
\newblock {\em EMS Surv. Math. Sci.}, 1(2):153--240, 2014.

\bibitem[Tur88]{Turaev}
V.~G. Turaev.
\newblock The {C}onway and {K}auffman modules of a solid torus.
\newblock {\em Zap. Nauchn. Sem. Leningrad. Otdel. Mat. Inst. Steklov. (LOMI)},
  167(Issled. Topol. 6):79--89, 190, 1988.

\bibitem[TV11]{TV11}
Bertrand To\"{e}n and Gabriele Vezzosi.
\newblock Alg\`ebres simpliciales {$S^1$}-\'{e}quivariantes, th\'{e}orie de de
  {R}ham et th\'{e}or\`emes {HKR} multiplicatifs.
\newblock {\em Compos. Math.}, 147(6):1979--2000, 2011.

\bibitem[vV17]{SV1}
Pavol \v{S}evera and Fridrich Valach.
\newblock Ricci flow, {C}ourant algebroids, and renormalization of
  {P}oisson-{L}ie {T}-duality.
\newblock {\em Lett. Math. Phys.}, 107(10):1823--1835, 2017.

\bibitem[vV20]{SV2}
Pavol \v{S}evera and Fridrich Valach.
\newblock Courant algebroids, {P}oisson-{L}ie {T}-duality, and type {II}
  supergravities.
\newblock {\em Comm. Math. Phys.}, 375(1):307--344, 2020.

\bibitem[Wil17]{Bri}
Brian Williams.
\newblock The {V}irasoro vertex algebra and factorization algebras on {R}iemann
  surfaces.
\newblock {\em Lett. Math. Phys.}, 107(12):2189--2237, 2017.

\bibitem[Wit87]{WitLoop}
Edward Witten.
\newblock Elliptic genera and quantum field theory.
\newblock {\em Comm. Math. Phys.}, 109(4):525--536, 1987.

\bibitem[Wit07]{Wit07}
Edward Witten.
\newblock Two-dimensional models with {$(0,2)$} supersymmetry: perturbative
  aspects.
\newblock {\em Adv. Theor. Math. Phys.}, 11(1):1--63, 2007.

\end{thebibliography}

\end{document}